\documentclass{article}
\usepackage[letterpaper,top=2cm,bottom=2cm,left=3cm,right=3cm,marginparwidth=1.75cm]{geometry}
\usepackage{graphicx} 
\usepackage{blindtext}
\usepackage{quantikz}
\usepackage{nicefrac}
\usepackage{algorithm}
\usepackage{algpseudocode}
\usepackage{amsmath}
\usepackage{amssymb}
\usepackage{siunitx}
\usepackage{array}
\usepackage{hyperref}
\usepackage{dsfont}
\usepackage{float}
\usepackage{subcaption}
\usepackage{authblk}
\usepackage{comment}
\usepackage{mathtools}
\usepackage{booktabs}
\usepackage[square,numbers]{natbib}
\hypersetup{colorlinks = true, citecolor=blue}

\def\corrcolor{black} 

\title{Exploring Quantum Annealing for Coarse-Grained Protein Folding}
\author[a,b]{Timon Scheiber\footnote{\href{mailto:timon.florian.scheiber@igd.fraunhofer.de}{timon.florian.scheiber@igd.fraunhofer.de}}}
\author[a,b]{Matthias Heller\footnote{\href{mailto:matthias.heller@igd.fraunhofer.de}{matthias.heller@igd.fraunhofer.de}}}
\author[a,b]{Andreas Giebel}
\affil[a]{Fraunhofer Institute for Computer Graphics Research IGD, Darmstadt, Germany}
\affil[b]{Technical University of Darmstadt, Interactive Graphics Systems Group, Darmstadt, Germany
}
\date{\today}

\begin{document}

\maketitle
\begin{abstract}
    We explore the potential application of quantum annealing to address the protein structure problem. To this end, we compare several proposed ab initio protein folding models for quantum computers and analyze their scaling and performance for classical and quantum heuristics. Moreover, we introduce a novel encoding of coordinate-based models on the tetrahedral lattice, based on interleaved grids. Our findings reveal significant variations in model performance, with one model yielding unphysical configurations within the feasible solution space. Furthermore, we conclude that current quantum annealing hardware is not yet suited for tackling problems beyond a proof-of-concept size, primarily due to challenges in the  embedding. Nonetheless, we observe a \textcolor{\corrcolor}{possible} scaling advantage over our in-house simulated annealing implementation, which, however, is only noticeable when comparing performance on the embedded problems.
\end{abstract}

\section{Introduction}
The prediction of a protein’s three-dimensional structure from its amino acid sequence has long been a central challenge in computational biology. Beyond the fundamental question of how a protein folds, a protein’s structure governs most of its interactions and is therefore critical for applications such as virtual ligand screening~\cite{wu2025folding} and the study of protein–protein interactions~\cite{keskin2008principles}, both of which are of great importance for modern \textit{in silico} pharmacology. 
Recent advances in artificial intelligence (AI) models have enabled the successful prediction of structures for a diverse array of proteins~\cite{jumper2021highly,baek2021accurate, abramson2024accurate}. However, structures  with no known homologues~\cite{doga2024perspective} or the estimation of  the physical folding pathway~\cite{outeiral2022current} still remain challenging.
Furthermore, incorporating non-canonical amino acids into the bioengineering process offers new opportunities and provides a dataset for which little training data exists \cite{birch2024noncanonical}.
In contrast, physics-based approaches, either simulating the physical folding process or searching for a conformation that minimizes a physics-inspired energy function, struggle with the immense conformational space and the rugged free-energy landscape characteristic of proteins~\cite{zhuravlev2010protein, trebst2006optimized}.
As a result, efforts to solve physics-inspired problems have increasingly focused on heuristic algorithms designed to efficiently navigate complex energy landscapes.
Finding the energy minimum of complex systems is a well-studied problem in statistical physics and combinatorial optimization. Naturally, several of the proposed heuristics to estimate the global minimum of a complex energy function have been adapted to the protein folding problem. Prominent examples include  simulated annealing~\cite{kirkpatrick1983optimization} and parallel tempering (which is often also denoted replica exchange method in the literature)~\cite{trebst2006optimized,swendsen1986replica, agostini2006generalized}.

The principal obstacle of this strategy is the rugged nature of the free-energy landscape: deep wells are separated by steep energy barriers, so gradient-based optimizers often stall in sub-optimal minima~\cite{zhuravlev2010protein}.  
In the software Rosetta~\cite{alford2017rosetta} this difficulty is partly alleviated by progressively ramping up the strength of repulsive terms, which helps the search to escape isolated wells.

An alternative route could be given by the advent of new quantum technologies, especially quantum annealing, an algorithm originally derived as a quantum analogue to simulated annealing~\cite{de1988numerical}.
By utilizing quantum tunneling, quantum annealing can potentially overcome energy barriers more rapidly than classical algorithms, accelerating the optimization process~\cite{kim2025quantum, denchev2016computational}. 
Since proteins possess notoriously rugged free-energy landscapes, quantum annealing could be pivotal for solving larger or novel protein structures, especially those with which AI-based models \textcolor{\corrcolor}{currently} struggle \cite{doga2024perspective}.

Perdomo-Ortiz et al.~\cite{perdomo2008construction} were the first to suggest a model to tackle the protein structure problem (PSP) on a quantum annealer. 
Due to the limitations of current hardware these models are restricted to coarse-grained models, folding the protein on a discrete lattice. Subsequent work has refined the approach through more efficient problem encodings~\cite{perdomo2012finding} or alternative lattice architectures \cite{robert2021resource}. This has given rise to multiple variants, each offering its own benefits and compromises~\cite{perdomo2008construction, perdomo2012finding,robert2021resource, babbush2014construction, babej2018coarse, irback2022folding, pamidimukkala2024protein}.
Apart from the PSP, recent studies have been performed to determine the feasibility of quantum computing approaches for both protein design~\cite{irback2024using} and protein–peptide docking~\cite{brubaker2024quadratic}.
However, current implementations have not yet escaped the proof-of-principle stage. In this work, we investigate the scaling of these models both in terms of their resource requirements and as the projected scaling of the time it takes to find the native fold. Our findings highlight that the correct choice of model is crucial in the noisy intermediate-scale quantum (NISQ) era and even beyond.

\paragraph{Related work and main contributions}
The application of quantum computing for protein folding has recently attracted significant attention due to its widespread applications. Despite current quantum computers not yet being capable of handling the complexities of relevant protein sizes, many studies have investigated their potential future use in this field. For example, Boulebane~\textit{et al.}~\cite{boulebnane2023peptide} investigated the potential of the quantum approximate optimization algorithm (QAOA) \cite{farhi2014quantum} for the protein structure problem, finding rather negative results in comparison to classical methods.
Further approaches to the problem using digitized counterdiabatic protocols as presented by Chandarana~\textit{et al.}~\cite{chandarana2023digitized} or Romero~\textit{et al.}~\cite{romero2025protein} which show more promising results than QAOA.
Outeiral~\textit{et al.}~\cite{outeiral2021investigating} explored the potential for limited quantum speedups by examining the scaling of the spectral gap for a dense problem encoding as the peptide chain length increases, finding exponentially quickly closing gaps for worst-case examples but only polynomially closing gaps in the average case. 
They also compared the performance of simulated annealing with ideal quantum annealing through direct numerical simulations of the Schrödinger equation for short peptide sequences. 
Furthermore, Linn~\textit{et al.}~\cite{linn2024resource} conducted a resource estimation for various approaches to the protein folding problem using gate-based quantum computers and QAOA.
Doga~\textit{et al.}~\cite{doga2024perspective} investigated the protein structure prediction problem with a focus on practical applications. 
They developed a framework to identify proteins that could benefit from quantum computing-based approaches, particularly those with a rugged free-energy landscape and limited homologues, to demonstrate an advantage over AI-based methods. Notably, they demonstrated that for a proof-of-principle protein (PDB: 5GJB) the quantum computing approach combined with classical post-processing could lead to lower root-mean-square errors at the all-atom resolution than AlphaFold2.

Our work builds on previous research by focusing specifically on the paradigm of quantum annealing.
To this end, we compare and revise several proposed formulations of the coarse-grained lattice protein folding problem in terms of their scaling in resource cost. We aim to identify which of these models could benefit from a quantum annealing approach by calculating the spin overlap distribution, a proxy for the complexity of the free-energy landscape. Finally, we compare the performance of classical heuristic solvers with quantum annealing hardware. 
To the best of our knowledge, our study is the first to perform a scaling analysis for multiple protein sequences on real, currently available quantum hardware and the first formulation-dependent comparison.

By performing benchmark calculations, we found that the model of Ref.~\cite{robert2021resource} produces non-physical folds with lowest energy, when the amino acid sequence is longer than $10$ residues, see Fig.~\ref{fig:UnphysicalConfig} and Appendix~\ref{sec:unphysical_fold} for more details.
Apart from this we introduce a novel encoding scheme for the PSP on quantum computers based on the works of Robert~\textit{et al.}~\cite{robert2021resource} in combination with the work of Babbush~\textit{et al.}~and Irbäck~\textit{et al.}~\cite{babbush2014construction, irback2022folding}. 
We find that for the shorter sequences considered in this work, our encoding provides the best observed performance. Furthermore, with the presented encoding we are able to embed larger sequences of up to $18$ amino acids onto current-gen quantum hardware. 
However, we were not able to solve them on the hardware using standard quantum annealing procedures.\\

The remainder of this article is structured as follows. 
In Section~\ref{sec:methods}, we provide the details of the considered methods, such as the models and solvers, used in this study. 
Section~\ref{sec:results} aims to establish the suitability of the considered problem formulations by investigating their resource scaling as well as the spin overlap distributions. 
In Section~\ref{sec:comp}, we compare the scaling of the time to solution for the best projected model between simulated annealing and quantum annealing. 
Finally, we conclude the study in Section~\ref{sec:conclusion}.

\section{Methods}
\label{sec:methods}
\subsection{Quadratic unconstrained binary optimization}

Quadratic Unconstrained Binary Optimization (QUBO) is the task of finding the minimal configuration for the problem
\begin{equation}
    \min_{b} \sum_{i,j} b_i Q_{ij} b_j,
\end{equation}
where \( b_i \in \{0,1\} \) are Boolean variables, and \( Q_{ij} \) is the real-valued QUBO matrix.

A closely related problem is the Ising spin glass problem from condensed matter physics. In this case, one seeks low-energy configurations for the generic Ising Hamiltonian
\begin{equation}
\label{eq:Ising}
    H_\text{Ising} = \sum_{i,j} J_{ij} s_i s_j + \sum_i h_i s_i,
\end{equation}
where the spin variables \( s_i \in \{-1, 1\} \) represent the two possible states of a spin. The Boolean variables \( b_i \) and the spin variables \( s_i \) are related through the linear transformation $b_i = \frac{1 + s_i}{2}$.
This transformation allows the Ising Hamiltonian to be directly mapped to a QUBO matrix, making the two formulations mathematically equivalent.
This relation between QUBOs and spin glasses triggered the development of several physics-inspired optimization algorithms, which originally were used to tackle many-particle problems in condensed matter physics.  
The most prominent examples, which we also use in this study, are simulated annealing (see Sec.~\ref{sec:SA}), quantum annealing (see Sec.~\ref{sec:QA}) and parallel tempering (see Sec.~\ref{sec:PT}).
Finding the ground state of Ising Hamiltonians is a notoriously difficult optimization problem and is generally known to be NP-hard~\cite{barahona1982computational}.
Throughout this manuscript, we will use the convention that \( s_i \) refers to variables in Ising space and \( b_i \) refers to variables in QUBO space.

In many applications the optimization problem that needs to be solved does not naturally take the form of a QUBO or Ising formulation and may contain higher-order terms
\begin{equation}
    \sum_i c_i b_i + \sum_{i,j} c_{ij} b_i b_j + \sum_{i,j,k} c_{ijk} b_i b_j b_k + ... \quad.
\end{equation}
For the later sections it will become important to map these higher-order unconstrained binary optimization (HUBO) problems to QUBO problems.
For example, in Boolean space, this can be accomplished using Rosenberg's polynomial~\cite{rosenberg1975reduction}, which can be used to reduce the order of a term by one at the cost of introducing additional variables. 
This way a term of degree three,
\begin{equation}
   H = b_1 b_2 b_3,
\end{equation}
can be transformed into a 2-local term, $H= b_1 b_4$, by introducing an auxiliary variable \( b_4 = b_2 b_3 \). To ensure that the auxiliary variable $b_4 = 1$, if and only if both \( b_2  = 1\) and \( b_3 =1 \), an additional penalty term needs to be added in the form of Rosenberg's polynomial
\begin{equation}
    H_\text{penalty} = \alpha (b_2 b_3 - 2 b_4 (b_2 + b_3) + 3 b_4).
\end{equation}
The strength $\alpha$ is crucial for the formulation and must be chosen sufficiently large to ensure that the original structure of the energy landscape is conserved. That is, there should not be a potential energy gain when the condition is not satisfied. By applying this method iteratively, any HUBO can be reduced to a QUBO at the cost of additional variables.

\subsection{Coarse-grained protein folding}
\begin{figure*}
    \label{fig:lower_error}
    \hspace{5em}
    \begin{subfigure}[h]{0.25\linewidth}
    \includegraphics[width=\linewidth]{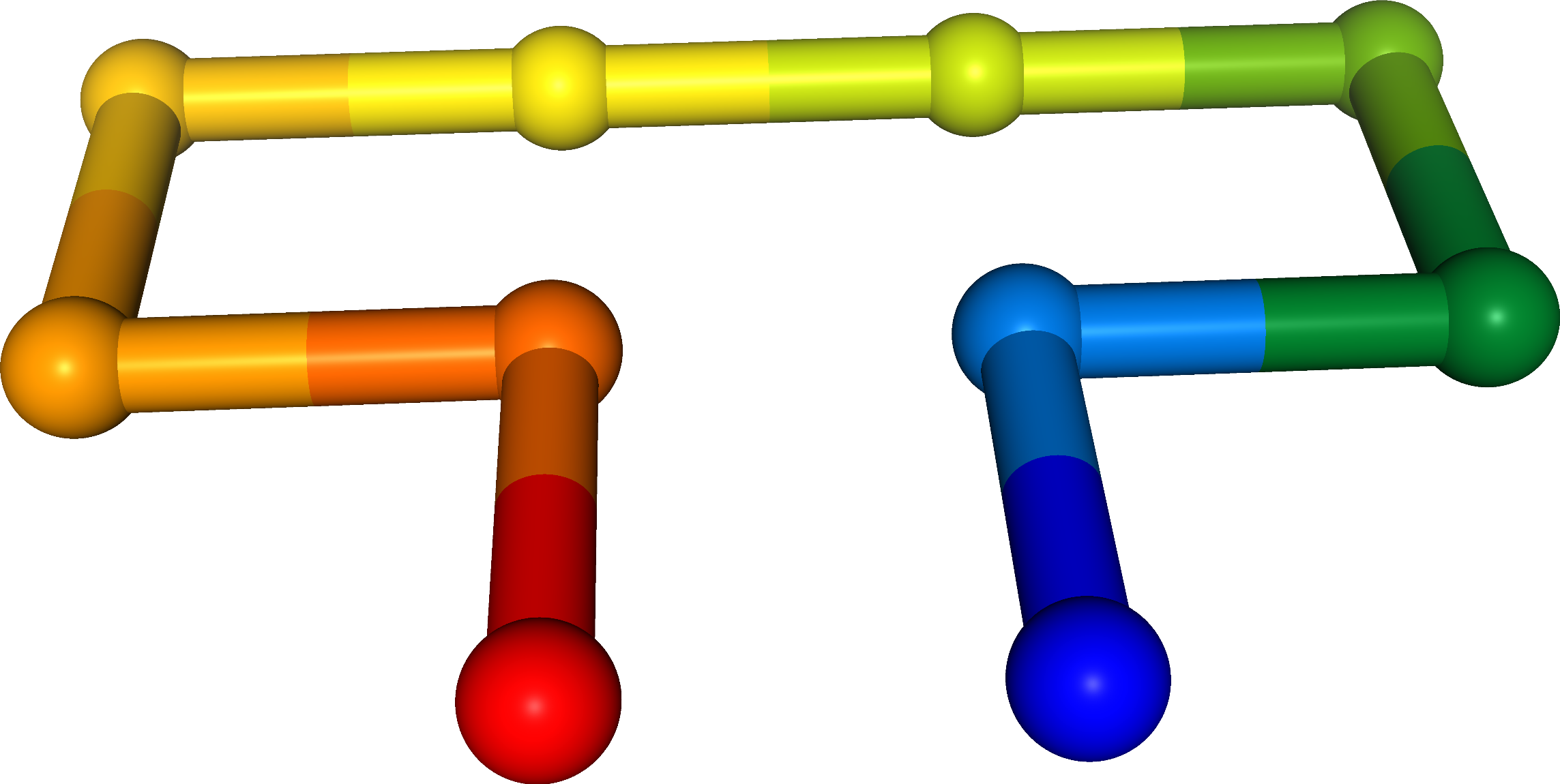}
    \caption{}
    \end{subfigure}
    \hspace{10em}
    \begin{subfigure}[h]{0.25\linewidth}
    \includegraphics[width=\linewidth]{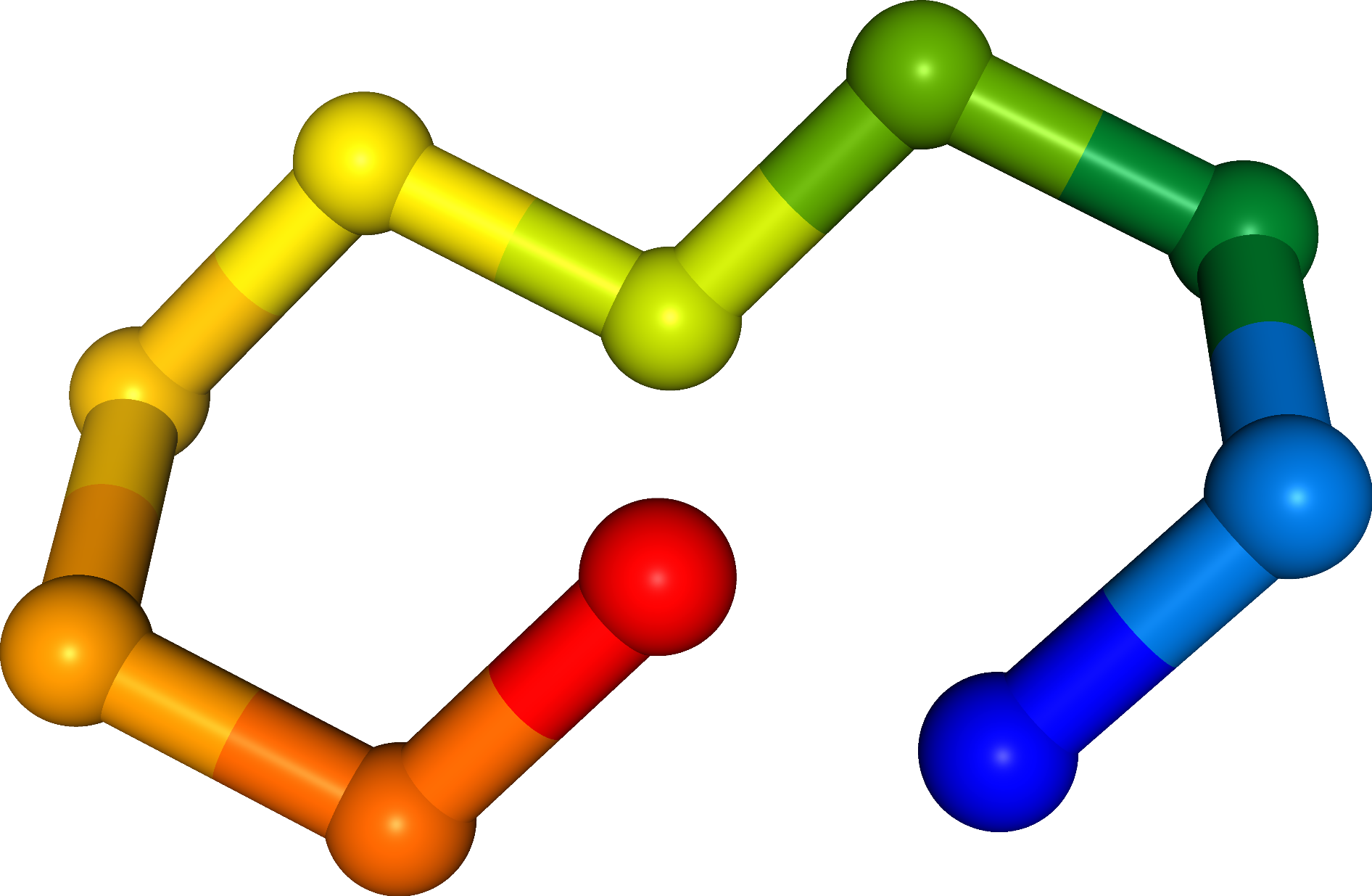}
    \caption{}
    \end{subfigure}%
    \caption{Example of a 10 amino acid mini protein folded on two different lattices. (a) A two-dimensional Cartesian grid and (b) A three-dimensional tetrahedral/diamond grid. Each amino acid corresponds to a single bead in the chain. Interactions between amino acids are established via nearest-neighbor interactions. Images were produced using the NGL viewer~\cite{rose2015ngl}.}
\label{fig:foldings}
\end{figure*}

To find the native fold of a given protein on current NISQ hardware, a problem formulation that adheres to the restrictions of the hardware must be adapted. 
The most commonly used approach involves formulating the problem on a coarse-grained lattice~\cite{perdomo2008construction, perdomo2012finding,robert2021resource, babbush2014construction, babej2018coarse, irback2022folding, pamidimukkala2024protein}. 
In this formulation, the protein is represented as a chain of multiple beads, where each bead corresponds to a single or multiple amino acids (see Fig.~\ref{fig:foldings}).

The positions a bead can take are discretized and interactions between amino acids are modeled according to nearest- (or next$^n$-nearest-) neighbor interactions on the lattice sites. 
The free energy of a given fold is derived from pairwise interactions of the amino acids either by a simple hydrophobic-hydrophilic (HP) model, the interaction matrix derived by Miyazawa and Jernigan~\cite{miyazawa1985estimation} or Lennard-Jones-type potentials~\cite{boulebnane2023peptide}.
Using these coarse-grained problem representations, finding the lowest energy configuration reduces to a discrete optimization problem (which can be mapped to a QUBO), which is even in the simplest HP case, known to be NP-complete~\cite{berger1998protein}.

Since the first formulation of the problem~\cite{perdomo2008construction} various model improvements have been made.
For example, two different ways of encoding the positions of the amino acids on the lattice have emerged; either direct encoding as coordinates \textcolor{\corrcolor}{on a finite size lattice of size $L$}, often denoted as a \textit{coordinate-based} models~\cite{perdomo2008construction}, or as a set of turns the polypeptide chain has taken, denoted \textit{turn-based} models \cite{perdomo2012finding}.

In this paper, we focus on the most promising near-future candidates that can be efficiently mapped onto a quantum annealer, specifically those models with bounded locality, that is models that have a limited number of qubits participating in an interaction. The models we investigate are:

\begin{itemize}
    \item[1.] A turn-based model on a three-dimensional Cartesian grid~\cite{babbush2014construction, babej2018coarse},
    \item[2.] a turn-based model on a tetrahedral grid~\cite{robert2021resource},
    \item[3.] a coordinate-based model on a Cartesian grid~\cite{babbush2014construction,irback2022folding}, and
    \item[4.] an adaptation of the coordinate-based model on the tetrahedral grid, which to the best of our knowledge has not yet been discussed in the literature.
\end{itemize}

\textcolor{\corrcolor}{The model proposed in this work adapts the coordinate-based encoding of Babbush~\textit{et al.}~\cite{babbush2014construction} and Irbäck~\textit{et al.}~\cite{irback2022folding} to a tetrahedral grid, as described in Robert~\textit{et al.}~\cite{robert2021resource}. In contrast to the other encodings presented, it aims to deliver a best-of-both-worlds approach: it preserves the native 2-local encoding of the coordinate-based model while leveraging a grid with a sparser interaction structure. 
The model maintains the underlying structure of the coordinate-based encoding and uses two interleaved grids derived from the face-centered cubic lattices, offset by a quarter diagonal to realize the tetrahedral grid. Compared with the approach of Brubaker~\textit{et al.}~\cite{brubaker2024quadratic}, our method does not enlarge the underlying grid and does not rely on penalty terms to enforce grid conformity, making it more resource-efficient. The multi-grid strategy could be readily extended to problems such as conformational docking, enabling efficient encoding of folding processes on two grids (e.g., a reaction pocket in a larger protein). More details on the encoding are presented in Appendix~\ref{Appendix:TetCart}.}
A review of all considered models, including some minor adjustments, can be found in Appendix~\ref{app:models}.
\textcolor{\corrcolor}{In Tab.~\ref{tab:model_comparision} we summarize the properties of the different models before the mapping to 2-local terms. Note, that this mapping, as described above, increases the number of qubits.}

\begin{table}[]
    \centering
    \begin{tabular}{|c|c|c|c|}
    \hline
        \text{Model} & \text{Qubit scaling} & \text{Locality}& \text{Introduced in}\\
        \hline
        \text{Turn-based Cartesian} &$\mathcal{O}(N^2 \log{N})$ & 8 (dense) / 4 (sparse) & Babej \textit{et al.}\cite{babej2018coarse} \\
        \text{Turn-based tetrahedral} & $\mathcal{O}(N^2)$ & 5 (dense) / 3 (sparse)&Robert \textit{et al.}~\cite{robert2021resource}\\
        \text{Coordinate-based Cartesian} & $\mathcal{O}(\nicefrac{N}{2}\cdot L)$ & 2& Irbäck \textit{et al.}~\cite{irback2022folding}\\
        \text{Coordinate-based tetrahedral} & $\mathcal{O}(\nicefrac{N}{2}\cdot (L_1 +L_2))$ & 2& this work\\
         \hline
    \end{tabular}
    \caption{\textcolor{\corrcolor}{Scaling properties of the models considered in this study, before the reduction to a 2-local model. Note that we do not extend the models beyond nearest-neighbor interactions in order to stay as resource efficient as possible. For the coordinate-based approaches the lattice sizes $L$ need to be scaled so that the whole sequence can fit onto the lattice $L > N$.}}
    \label{tab:model_comparision}
\end{table}

\subsection{Simulated annealing}
\label{sec:SA}
Simulated annealing (SA) is a meta-heuristic algorithm that has been adapted from \textcolor{\corrcolor}{statistical mechanics} to the field of optimization~\cite{kirkpatrick1983optimization}.
The algorithm employs a temperature-based Markov chain Monte Carlo (MCMC) method to sample low-energy states, mimicking the physical annealing process of metals. 
Its primary advantage over naive Monte Carlo approaches lies in efficient sampling through the Metropolis criterion~\cite{metropolis1953equation}.
Starting from an initial state with energy \(E_\text{curr}\), a new state with energy \(E_\text{prop}\) is proposed. If the energy of the proposed state is lower than that of the current state, the transition is accepted. If not, the transition is accepted with a probability given by
\begin{equation}
    p = e^{-\beta \Delta E},\label{eq:sa_prob}
\end{equation}
where \(\beta = \frac{1}{kT}\) denotes the inverse temperature\footnote{The usage of $k$ has a physical context and ensures equal units for temperature and energy. For the simulation we utilize units of $k=1$.} and \(\Delta E = E_\text{prop} - E_\text{curr}\) is the energy difference. This probability allows the algorithm to escape from local minima where it might otherwise become trapped. To ensure convergence to a minimum, the temperature $T$ is lowered with an exponential cooling schedule at a selected cooling rate $\zeta \in (0, 1)$, \textcolor{\corrcolor}{such that after an attempt to flip each spin the temperature is reduced according to} \begin{equation}
    T_{i+1} = \zeta \cdot T_i.
\end{equation}

In our implementation, the start temperature $T_0$ is automatically selected based on $n$ random spin flips performed on a random state vector with the following algorithm proposed by Atiqullah \cite{Atiqullah.2004}:
\begin{equation}
    T_0=\frac{\overline{\Delta E} + 3 s_{\Delta E}}{\ln{\left(\frac{1}{\chi}\right)}} \quad \text{with } \quad \chi = \frac{n_{\mathrm{flipped}}}{n},
\end{equation}
where $n_{\mathrm{flipped}}$ is the number of accepted spin flips for $n$ tries, 
$\overline{\Delta E}$ represents the sample mean and $s_{\Delta E}$ being the sample standard deviation of the energy difference per flip.

For parallelization, the algorithm employs a multi-flip procedure by performing the spin flips on each independent set of the QUBO matrix graph in parallel, similar to the method described in Ref.~\cite{Imanaga2021} by Imanaga \textit{et al}. 
The main difference is that we apply the Deveci graph coloring heuristic \cite{Deveci2016ParallelGC} to determine independent sets in the QUBO graph. 
The sparser the graph, the more the algorithm can use GPU parallelism, usually resulting in more independent nodes per set and therefore a higher parallelization potential.

Since usually a single run of the algorithm will not return the global energy minimum, the algorithm is repeated several times to sample a distribution of low energy states.
To speed up the sampling, we use a GPU-parallelized SA implementation running on two NVIDIA A100 GPUs. This setup enables the sampling of 432 separate instances in parallel across all considered problem sizes. We want to highlight that the implicit parallelization speedup is considered in all results of this manuscript.

\subsection{Quantum annealing}
\label{sec:QA}
Simulated quantum annealing, as introduced in Refs.~\cite{kadowaki1998quantum,doi:10.1126/science.1068774}, is a quantum-inspired algorithm to solve combinatorial optimization problems, that runs on classical hardware.
The actual physical implementation, i.e., quantum annealing (QA) on dedicated hardware, is a non-universal form of quantum computing aimed at solving combinatorial optimization (CO) problems that are classically difficult to tackle.

Quantum annealers solve optimization problems (quasi-)adiabatically by initializing an easy-to-prepare ground state and gradually ramping up a problem Hamiltonian while ramping down the initial Hamiltonian. 
The Hamiltonian can be written as
\begin{equation}
    \hat{H}_\text{QA}(s) = A(s) \hat{H}_\text{initial} + B(s) \hat{H}_\text{problem},
\end{equation}
where \( s = t/t_a \) is a dimensionless parameter that characterizes the Hamiltonian at each time $t$ during the annealing process, with maximal time $t_a$.

The functions \( A(s) \) and \( B(s) \) are amplitudes that typically satisfy the boundary conditions \( A(0) \gg B(0) \) and \( B(1) \gg A(1) \), ensuring that at the end of the annealing process, only the problem Hamiltonian contributes to the energy landscape. 
Unlike the broader concept of adiabatic quantum computing, QA only realizes stoquastic Hamiltonians~\cite{bravyi2006complexity}, making it a non-universal form of quantum computing. 
In current hardware, the problem Hamiltonian \( \hat{H}_\text{problem} \) is encoded as the Ising Hamiltonian 
\begin{equation}
\label{equation:Ising}
    \hat{H}_\text{Ising} = \sum_{i,j} J_{ij} \hat{Z}_i \hat{Z}_j + \sum_i h_i \hat{Z}_i,
\end{equation}
which consists of the programmable parameters \( J_{ij} \), denoting the inter-qubit couplings, as well as the single qubit biases \( h_i \).

\subsection{Parallel tempering and problem hardness}
\label{sec:PT}
Parallel tempering (also known as replica exchange Monte Carlo) is another temperature‐based heuristic for locating low‐energy configurations in an Ising spin glass. 
Unlike simulated annealing, which cools a single system along a predefined schedule, parallel tempering runs multiple copies (replicas) of the system in parallel at fixed temperatures $T_1 < T_2 < \dots < T_M$.
However, for each replica, the Monte Carlo sweeps are performed using the same spin flip acceptance probability as described in Eq.~\eqref{eq:sa_prob} for simulated annealing, at the corresponding replica temperature. 
Additionally, after sweeping through all spins in all replicas, one performs a swap of the assigned temperatures between two neighboring replicas (i.e., replicas with close temperatures) with the probability 
\begin{equation}
    p_{\rm swap} = e^{(E-E') (1/kT-1/kT')},\label{eq:replica_exchange}
\end{equation}
where $E$, $E'$ and $T$, $T'$ are the energies and temperatures of the two replicas, respectively.
Parallel tempering is parallelized in the same way based on graph coloring as described in Sec.~\ref{sec:SA} for SA.

We use parallel tempering to estimate the hardness of the different protein folding models by scanning the energy landscape for local minima.
Specifically, to quantify the usefulness of utilizing QA for a given problem, we use the order parameter
\begin{equation}
 q = \frac{1}{N}\sum_{i} \langle s_i^{(1)} s_i^{(2)} \rangle
\end{equation}
from spin-glass theory\footnote{\textcolor{\corrcolor}{In some parts of this work the letter $q$ is utilized to denote qubits. The meaning should be clear from context.}} as discussed in Refs.~\cite{yucesoy2013correlations,katzgraber2015seeking}. The index \( i \) runs over the individual spins of the problem, while the superscripts denote the index of two replicas of the problem with the same disorder, i.e., at the same temperature.
Thus, to calculate $q$, we have to run two \textcolor{\corrcolor}{independent} parallel tempering instances in parallel.

The order parameter serves as a measure of the average thickness of the barriers between local minima in the energy landscape. 
If the problem has many local minima, which can be reached from one another with only a few spin flips, the probability $P(q)$ of measuring an overlap $ q \approx 1$  is high. In this scenario, Ref.~\cite{katzgraber2015seeking} argues that quantum annealing has an advantage, as the quantum tunneling effect can help the optimizer to jump between minima. 
Conversely, if local minima are far from each other, i.e., many flips are required to jump from one to another, the spin overlap is closer to $0$. In this case, the problem has thick barriers and is notoriously difficult to solve, both for classical algorithms and quantum annealing.

The distribution $P(q)$ of the order parameter is especially interesting for the PSP since it can be interpreted as a proxy for the free-energy landscape \cite{yucesoy2013correlations}.

\section{Resource estimation for quantum annealing}
\label{sec:results}
In this section we investigate the scaling of the different protein folding models.
We first \textcolor{\corrcolor}{compare} different metrics like the number of qubits a quantum annealer needs to run these models, the density and the required resolution of the couplers (Sec.~\ref{sec:scaling}).
We then investigate the structure of the free-energy landscape to see if the models are amenable to quantum speedup due to quantum tunneling (Sec.~\ref{sec:spin_ovlp}).
Finally, we investigate the influence of the embedding process on the models (Sec.~\ref{sec:emb}).
In Sec.~\ref{sec:disc} we give a short summary and discussion of all results.

\subsection{Model scaling}
\label{sec:scaling}
\begin{figure}[!htb]
  \includegraphics[width=1\linewidth]{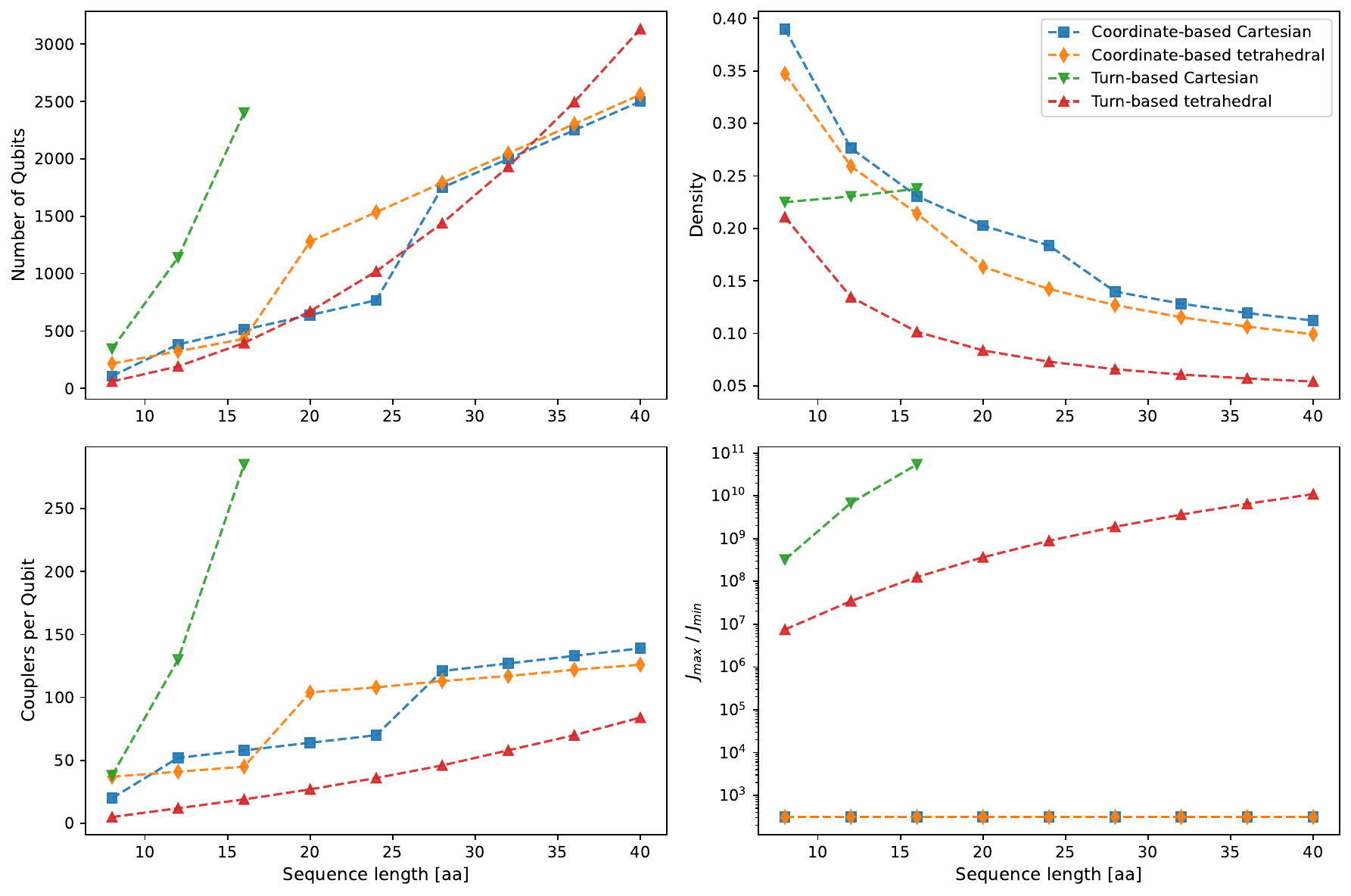}
  \caption{Scaling of the considered metrics. We show the required number of qubits when reducing the problem to a two-local QUBO (top left), the density of the QUBO matrix \( Q \) (top right), the average required number of qubit-qubit couplings per qubit (bottom left) and the maximum coupling strength divided by minimum coupling strength (bottom right). 
  All results depict the model metrics before embedding onto a given hardware graph. The stepwise increases in the coordinate-based models indicate points at which the grid size was adjusted.
  We investigated the turn-based Cartesian model only up to a sequence length of $16$ amino acids, since the reduction to a two-local model became too time-consuming for larger sequences.}
  \label{fig:scaling}
\end{figure}
To investigate the scaling properties we generate QUBO instances for each model ranging from $N=8$ to $N=40$ amino acids. 
Since the D-Wave devices only support 2-local couplings, we reduce the locality of all HUBOs (turn-based models) using Rosenberg's polynomial via the PyQUBO library~\cite{zaman2021pyqubo}.
As mentioned earlier, we have to choose the penalty strength for these additional variables to ensure that Rosenberg's polynomial conserves the energy landscape of the original problem. 
For our analysis we consider
\begin{equation}
    \alpha = 1 + \sum_i |c_i| + \sum_{i,j} |c_{ij}| + \sum_{i,j,k} |c_{ijk}| + ...
\end{equation}
which is a worst-case scaling, to ensure that the ancilla variables obey the constraints.
In the general case this leads to large QUBO coefficients which can be detrimental to performance. 
For example, a better choice could be made following the ideas of Ref.~\cite{babbush2013resource}.
However, the general scaling of the magnitudes of the coefficients with the sequence length will still remain.

We evaluate which model most effectively maps onto current-gen quantum annealers based on the scaling of various relevant properties of the models. 
These properties include the required number of logical qubits\footnote{By logical qubits we refer to the numbers of physical qubits needed if the device would support all-to-all connectivity.},  the density of the QUBO matrix, the average number of required couplers per qubit, and the minimal required coupler resolution \( J_\text{max}/J_\text{min} \). 
The results are presented in Fig.~\ref{fig:scaling}. Due to a steep increase in the computational time we only consider the turn-based model up to approximately $16$ amino acids. Beyond this sequence length we find that the generation of the QUBO matrix, especially regarding the reduction to a 2-local model, takes too long to be considered feasible. \textcolor{\corrcolor}{We find that for the execution on a 2-local quantum annealer this effect alone could preclude any possible quantum advantage for the turn-based Cartesian model.}

The coordinate-based model is defined on a finite grid with $L_\text{total} = L_x L_y L_z$ lattice sites (see Appendix~\ref{Appendix:CoordinateBased}), requiring the grid size to be specified prior to generating the QUBO matrix. For simplicity, we limit our analysis to symmetric grids where $L_x = L_y = L_z = L$. However, in certain scenarios, asymmetric grids (where $L_i \neq L_j$) may be more advantageous. Further, to maximize resource efficiency, we restrict our analysis to the minimal lattice size.
Since the size of the native fold (i.e., the minimum number of lattice sites needed to accommodate it) is not known a priori, we start with the smallest lattice capable of supporting the entire sequence, $L^3 \approx N$. As this is often too restrictive, we increment the grid length $L$ by 1 to provide additional degrees of freedom. For a Cartesian lattice, this corresponds to a minimal grid size of $L_\text{min} = \lceil N^{\frac{1}{3}} \rceil + 1$, while for the tetrahedral lattice, it is $L_\text{min} = \lceil (N/2)^{\frac{1}{3}} \rceil + 1$.

For the considered parameters we find a roughly equivalent scaling in the number of qubits for three out of the four models with the turn-based model on the Cartesian grid being the outlier.
Conversely, the turn-based model on the tetrahedral grid performs surprisingly well, even considering the additional resources that are required for the mapping of higher-order terms to $2$-local terms.

The density $\rho$ of the QUBO matrix relates to the number of qubit-qubit interactions required, relative to the maximum possible number of interactions.
Generally, it is conjectured that QA performs best for QUBOs with low density \cite{kim2025quantum}. Our findings show that, across all considered models, apart from the turn-based model on the Cartesian grid, the density decreases as the number of amino acids increases.
Additionally, the data reveals that the  turn-based model on the tetrahedral grid yields the sparsest QUBO matrix, making it potentially more suitable for quantum annealing.

Unlike QUBO density, the average number of couplers per qubit directly reflects the connectivity a device needs, to host the models without embedding. For every model studied, this value increases with sequence length, indicating a corresponding rise in embedding overhead.

Finally, we investigate the required coupler resolution \( J_\text{max}/J_\text{min} \) for each of the models, which is given by the absolute value of the quotient of the largest programmable coupler strength in relation to the lowest non-zero coupler strength. 
We find that, while the resolution is constant for the coordinate-based models, the resolution needs to be increasingly high for the turn-based models. 
As already studied in Ref.~\cite{babbush2014construction}, this effect is mostly induced by the reduction to a 2-local model, \textcolor{\corrcolor}{which increases the couplings due to the multiplication of penalty terms.}

\subsection{Spin overlap distributions}
\label{sec:spin_ovlp}
To determine if the given problems are suitable for quantum annealers, we estimate the distribution of the order parameter $q$ or spin overlap distribution (SOD) $P(q)$ for each problem formulation. 
To improve the simulation we chose the penalty terms/value of $\alpha$ for the turn-based models lower than in the scaling analysis. We provide details on the chosen penalties in Appendix~\ref{Appendix:TetCart}.
For the considered coordinate-based models we chose a constant grid size, consisting of $4^3 = 64$ lattice sites for the Cartesian grid as well as $2\cdot 3^3 =  54$ sites for the tetrahedral grid.
The SOD is estimated using parallel tempering as described by Katzgraber \textit{et al.}~\cite{katzgraber2015seeking}. 
We calculate the spin overlap from two parallel runs of parallel tempering using $ N_\text{steps}=6\cdot 10^6$  Monte Carlo sweeps, with 400 different temperature instances distributed geometrically between $T_\text{min}$ and $T_\text{max}$ \textcolor{\corrcolor}{(see Appendix \ref{Appendix:Supp_ParallelTemp} for the concrete values)}. The overlap distribution $P(q)$ is estimated by computing the spin overlap over $N_\text{olap} $ sweeps, which are performed after an initial thermalization period of $ N_\text{steps} - N_\text{olap} = 5\cdot 10^6$. 
This thermalization period ensures convergence to a local minimum for the lowest temperature instances.

The spin overlap is then extracted from the replicas corresponding to the lowest temperature of both instances. More details of the PT parameter choices for all simulations are provided in Appendix~\ref{Appendix:Supp_ParallelTemp}.
\begin{figure}[!htb]
    \centering
    \includegraphics[width=\linewidth]{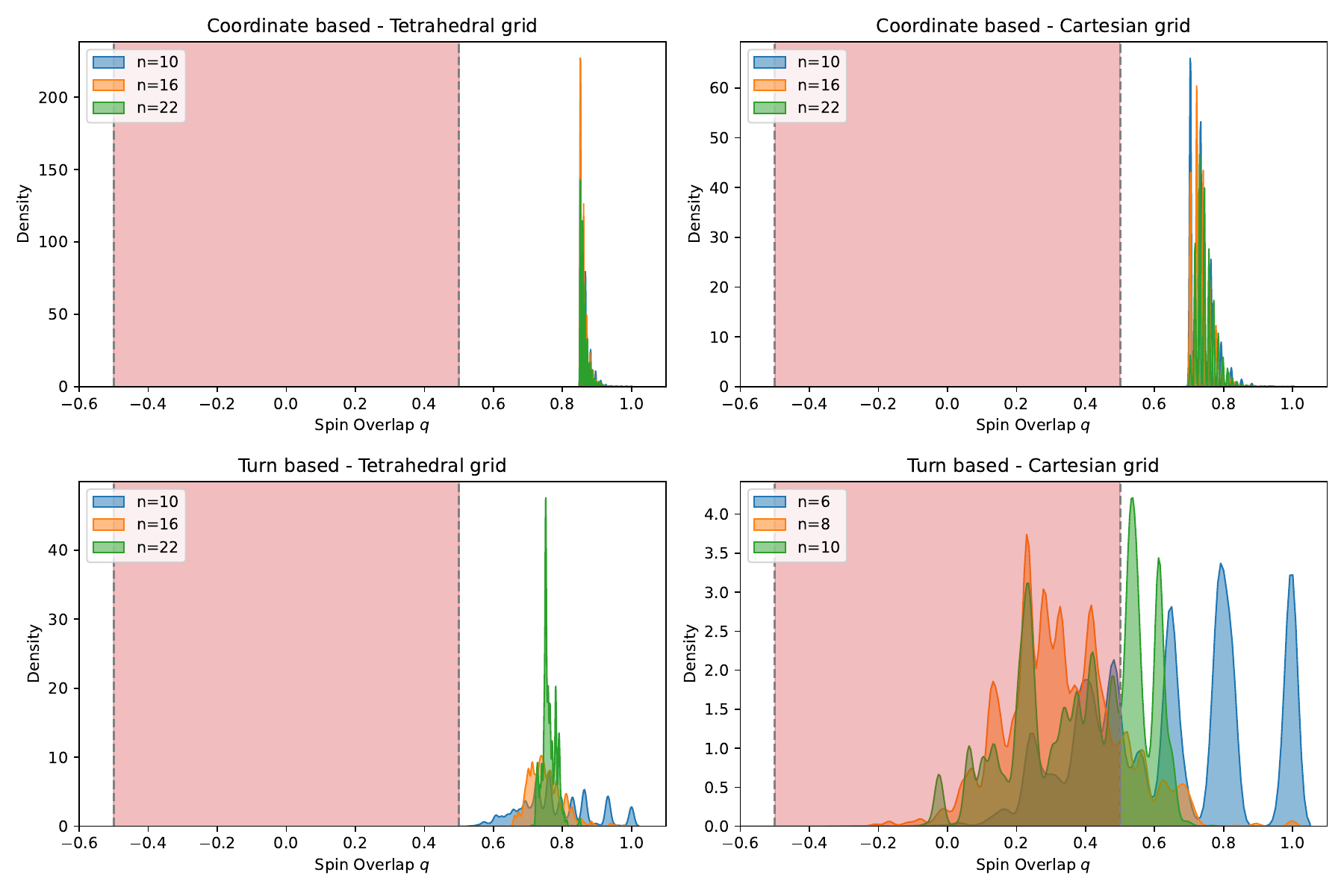}
    \caption{Spin overlap distribution for sections of increasing sequence length of the 189 amino acid protein M-RAS. The choice of protein is arbitrary and serves as a mere indication of the model differences for the same protein. 
    The results show the SOD for the coordinate-based model on the  tetrahedral (top left panel) and Cartesian grid (top right panel), as well as the turn-based models on the tetrahedral (bottom left panel) and  Cartesian grid (bottom right panel). 
    Areas highlighted in red indicate the range of thick barriers, where no quantum speedup due to quantum tunneling is expected~\cite{katzgraber2015seeking}.}
\label{fig:spin_overlap}    

\end{figure}

To see how the SOD evolves with growing sequence length we investigate growing sections of increasing sequence length of the M-Ras protein (PDB ID : 9C1A) for three discrete values of $10$, $16$ and $22$ amino acids. Since the turn-based model on the Cartesian grid required a substantially larger amount of resources, both in compute time as well as QUBO matrix size we restrict the SODs for this model to $6$, $8$ and $10$ amino acids. 
The results of the estimated spin overlap distributions are presented in Fig.~\ref{fig:spin_overlap}. As highlighted by the data the overlap distributions take vastly different forms for the considered models even though they encode the same protein.

We briefly analyze the measured SODs for the various models. 
All SODs lie predominantly in the regime $|q|> 0.5$. This results from the fact that for the lowest-temperature replica, the system relaxes into a local minimum that satisfies the penalty terms of the original formulation. Depending on the structure of the problem the overlap between two configurations that obey the constraints is generally larger.
As shown in Fig.~\ref{fig:spin_overlap}, the coordinate-based encodings produce a sharply peaked, discrete SOD. This arises from their one-hot encoded structure where each solution vector is partitioned into blocks in which exactly one spin is in the $+1$ state while the rest are in the $-1$ state.
The overlap between two such blocks can therefore take only a few discrete values, corresponding to either all spins aligned or at most two spins counter aligned, yielding the observed discrete spikes.

The turn-based encodings display a broader, less-structured SOD. 
Although the turn variables also discretize the landscape, additional qubits (like the interaction qubits or those introduced by Rosenberg’s polynomial) are subject to less structured penalties. 
As a result of these additional degrees of freedom, the overlap spectrum flattens into the diffuse profile observed.

Following the definitions of Ref.~\cite{katzgraber2015seeking} we evaluate the hardness of instances by considering the distribution of measured peaks in the SOD where we consider instances with all peaks in the regime $\vert q \vert > 0.5$ as instances with \textit{thin} barriers and instances with peaks outside this regime as instances with \textit{thick} barriers.
The region of thick barriers is indicated as a red shaded region.

The measured SODs for all but the turn-based model on the Cartesian grid lie in the regime of thin barriers where most peaks are located at $\vert q \vert > 0.5$. It is important to note that this effect stems in part from the fact that we chose a denser encoding for this model, as explained in Appendix~\ref{Appendix:TBCart}. 
The coordinate-based models clearly exhibit the most rugged energy landscape, as indicated by the closely spaced peaks. This suggests that the coordinate-based formulation is more suitable for leveraging quantum advantage through tunneling effects compared to the turn-based models.

\subsection{Embeddings}
\label{sec:emb}
A further restriction of currently available quantum annealers is the limited connectivity of the qubits. 
In physical systems, not all two-local interactions \( J_{ij} \) can be set because some qubits do not share a physical coupling. 
To solve problems requiring interactions between qubits not present in the hardware connection graph, an additional step called \textit{minor-embedding} must be utilized~\cite{choi2008minor}. The (minor-) embedding process involves finding a mapping from a given problem graph to the hardware graph by allowing for the contraction and removal of edges from the hardware graph until it matches the problem graph. 
While this allows for the solution of denser problems, it comes at the cost of an increased number of qubits as a chain of several physical qubits encode a single logical qubit. To ensure that all qubits in the chain are in the same state, the qubits are coupled ferromagnetically with a tunable \textit{chain strength}. The correct choice of this chain strength can generally have a large impact on the solver performance.

Finding a graph minor is NP-hard when the goal is to minimize the number of nodes~\cite{gomez2025addressing}. 
Consequently, practical applications rely on heuristics such as D-Wave’s minor-embedding algorithm, \textit{MinorMiner}~\cite{cai2014practical}.

\begin{figure}[!htb]

  \includegraphics[width=\linewidth]{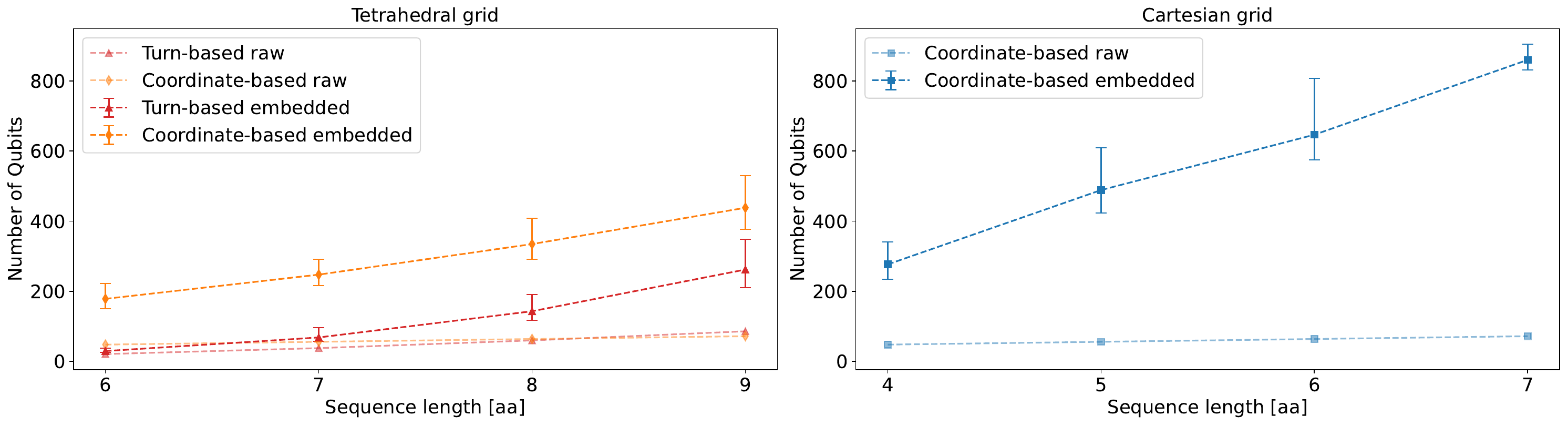}
  
  \caption{Required number of Qubits after the embedding for different sequence lengths and models. The data was taken for 1000 calculated embeddings, the error bars indicate best and worst case instances. Even for these short sequence lengths the embeddings can vary by more than a hundred qubits.}
  \label{fig:embedding}
\end{figure}

A key advantage of the considered models is their uniform structure across all proteins, with only the QUBO coefficients varying. 
This enables the reuse of embeddings, allowing a single efficient embedding to be applied to all proteins of the same size. 

To investigate the effect of the embedding for the different formulations, we generate 1000 embeddings for each protein size using D-Wave's \textit{minor-miner} for the \textit{Advantage 2 prototype}.
The \textit{Advantage 2 prototype} is an annealer with roughly $1200$ qubits based on the so-called \textit{Zephyr} topology \cite{dwave_prototype,king2022zephyr}.

Due to the device restrictions, we focus the analysis on shorter sequences, ranging from $6$ to $9$ amino acids for the tetrahedral grid and $4$ to $7$ amino acids for the Cartesian grid. We specifically chose this range as 4 (6) is the minimal sequence length to establish a nearest-neighbor contact between two amino acids on the Cartesian (tetrahedral) grid. All data regarding the coordinate-based models are taken with respect to the minimal grid that supports the native fold, as we found that after increasing the grid size we were not able to find a valid embedding. 

Due to the steep resource costs we omit the turn-based model on the Cartesian grid from the embedding analysis.
The scaling of the embeddings for the \textit{Advantage 2 prototype} are presented in Fig.~\ref{fig:embedding}. As shown the embedding greatly increases the resource cost for all models. Generally we found that sparser models require fewer physical qubits after the embedding. 

The error bars indicate the range between the worst and best case instances in the number of qubits. Some additional information regarding the distribution of the embeddings is presented in Appendix~\ref{Appendix:QA}.

\begin{figure}[!htb]
  \includegraphics[width=\textwidth]{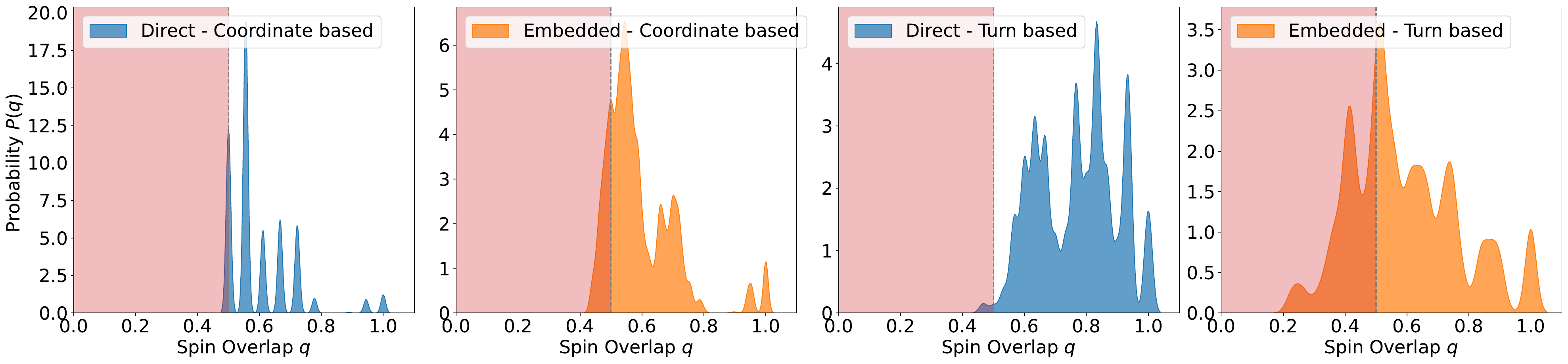}
  \caption{Influence of the embedding on the spin overlap for the models on the tetrahedral grid for an example of a protein with sequence length $7$. 
  The dashed line indicates a spin overlap of 0.5 as specified in Ref.~\cite{katzgraber2015seeking}. As shown, the embedding process leads to an increase in thickness of the energy barriers for the turn-based model. For the coordinate-based model this effect seems less pronounced.}
  \label{fig:Embed_Energy_Landscape}
\end{figure}

A further problem of the minor embedding process is that typically the embedded problem is more complex to solve in contrast to the direct problem due to the larger required number of qubits. 
To investigate if the embedding has an effect on the SOD and thus the ruggedness of the free-energy landscape, we embed a protein consisting of $7$ amino acids using the coordinate-based as well as turn-based encodings on the \textit{Zephyr} graph \cite{king2022zephyr} as an exemplary test case. 
We chose the chain strength of the embedding as half the largest (absolute) value of the QUBO matrix. 
We found that this choice in chain strength conserves the ground state energies while leading to improved performance in comparison to unnecessarily larger\ chain strengths.

The results are presented in Fig.~\ref{fig:Embed_Energy_Landscape} and show the influence of the embedding on the SOD $P(q)$ for the coordinate as well as turn-based models on the tetrahedral grid. 
Our findings highlight a general broadening of the measured SOD compared to the original problem.
The increase in degrees of freedom seems to affect the SOD which can in some cases shift the model to the area associated with thicker energy barriers.

\subsection{Discussion}
\label{sec:disc}
To conclude this section, we discuss the obtained results and evaluate whether the problem, in its current form, is suitable for a quantum annealing approach using D-Wave's sparsely connected hardware.
During the tests, we thoroughly investigated the proposed models beyond the regimes in which they were initially tested. We identified several flaws in the models that may prohibit their use with quantum annealers. Below, we provide a brief review of the main drawbacks of each of the tested models.

\paragraph{Turn-based Cartesian}
Throughout our analysis, we found that the turn-based model on the Cartesian grid performed the worst across nearly all considered metrics.
We discovered that mapping the model to a 2-local Hamiltonian requires a large number of auxiliary qubits and results in a dense QUBO matrix, further increasing the qubit count in the embedding. Additionally, we noted that the coupler resolution increases with problem size, requiring several orders of magnitude in resolution. As highlighted in Ref.~\cite{babbush2014construction}, this large resolution is a consequence of reducing the 12-local model to a 2-local one.
The drawback of the required resolution is twofold. First, classical (temperature-based) optimizers often struggle to traverse steep energy barriers. While this issue can be mitigated by selecting sufficiently high temperatures, other terms (such as the MJ interaction energies) have much lower magnitudes, meaning the height of these barriers becomes significant only in the later stages when the temperature is sufficiently low, hence making it difficult to explore new folds while also optimizing their energy.

The second drawback arises from the fact that couplers in a D-Wave device are affected by integrated control errors (ICE). 
These errors indicate that a coupler \(J_{ij}\) can only be set with some integrated error \(\delta J_{ij}\). 
Such errors can significantly degrade the performance of the quantum annealing approach, an effect known as J-chaos~\cite{pearson2019analog}. 
Especially for problems which require a resolution beyond the magnitude of these errors, they can be detrimental for performance.

\paragraph{Turn-based tetrahedral}
We found that the turn-based tetrahedral model performs surprisingly well across all considered metrics. 
Although originally proposed for use with a gate-based quantum computer, the derived 2-local models result in comparable qubit counts to the natively 2-local coordinate-based models, while being considerably sparser.
\textcolor{\corrcolor}{
However, due to the necessary scaling of the penalty terms, the model shares the same drawback of requiring high coupler resolution, which can limit performance on both quantum annealers and classical temperature-based solvers.}\\

\textcolor{\corrcolor}{Given the vastly different performance, let us summarize the improvements of the turn-based tetrahedral model over the turn-based Cartesian model. For a detailed discussion of the models, refer to Appendix~\ref{app:models}.
First, the grid change reduces the number of possible turns from six directions on the Cartesian grid to only four on the tetrahedral grid. Consequently, the number of conformation qubits per chain length $N$ changes from $6N$ (sparse encoding) or $\lceil \log_2 6 \rceil N = 3N$ (dense encoding) to $4N$ (sparse encoding) or $\log_2 4\,N = 2N$ (dense encoding).
In addition, the grid structure is considerably sparser, so fewer possible amino acid interactions have to be considered on the tetrahedral grid compared with the Cartesian grid.
Apart from these grid-related advantages, improvements also arise from the penalty terms used in Ref.~\cite{robert2021resource} to exclude overlapping folds. In the Cartesian model from Refs.~\cite{babbush2014construction, babej2018coarse}, for each pair of beads $j$ and $k$ that could overlap, a (squared) distance function $D(j,k)$ is introduced to ensure that only configurations with $D(j,k)>0$ are feasible. The construction of these penalty terms is non-trivial and requires auxiliary qubits (slack variables) to transform the inequality into an equality (see Appendix~\ref{Appendix:TBCart}).}

\textcolor{\corrcolor}{
In contrast, the tetrahedral model penalizes overlaps only in the direct vicinity of possible nearest-neighbor interactions. This local formulation allows the overlap constraints to be absorbed into the energy function, avoiding the need for additional qubits in the form of slack variables.
Lastly, the turn-based Cartesian model generally has higher locality than the turn-based tetrahedral model (see Appendix~\ref{Appendix:TBtet} and Sec.~\ref{sec:methods}). Since higher locality typically increases the qubit overhead via the reduction to a 2-local model, the reduced locality of the turn-based tetrahedral approach contributes to its improved performance relative to the turn-based Cartesian one.
Together, these effects explain the superior performance of the tetrahedral model over the Cartesian formulation.
}

\begin{figure}[!htb]
\hspace{1cm}
  \begin{subfigure}{.3\linewidth}
  \centering
    \includegraphics[width=\linewidth]{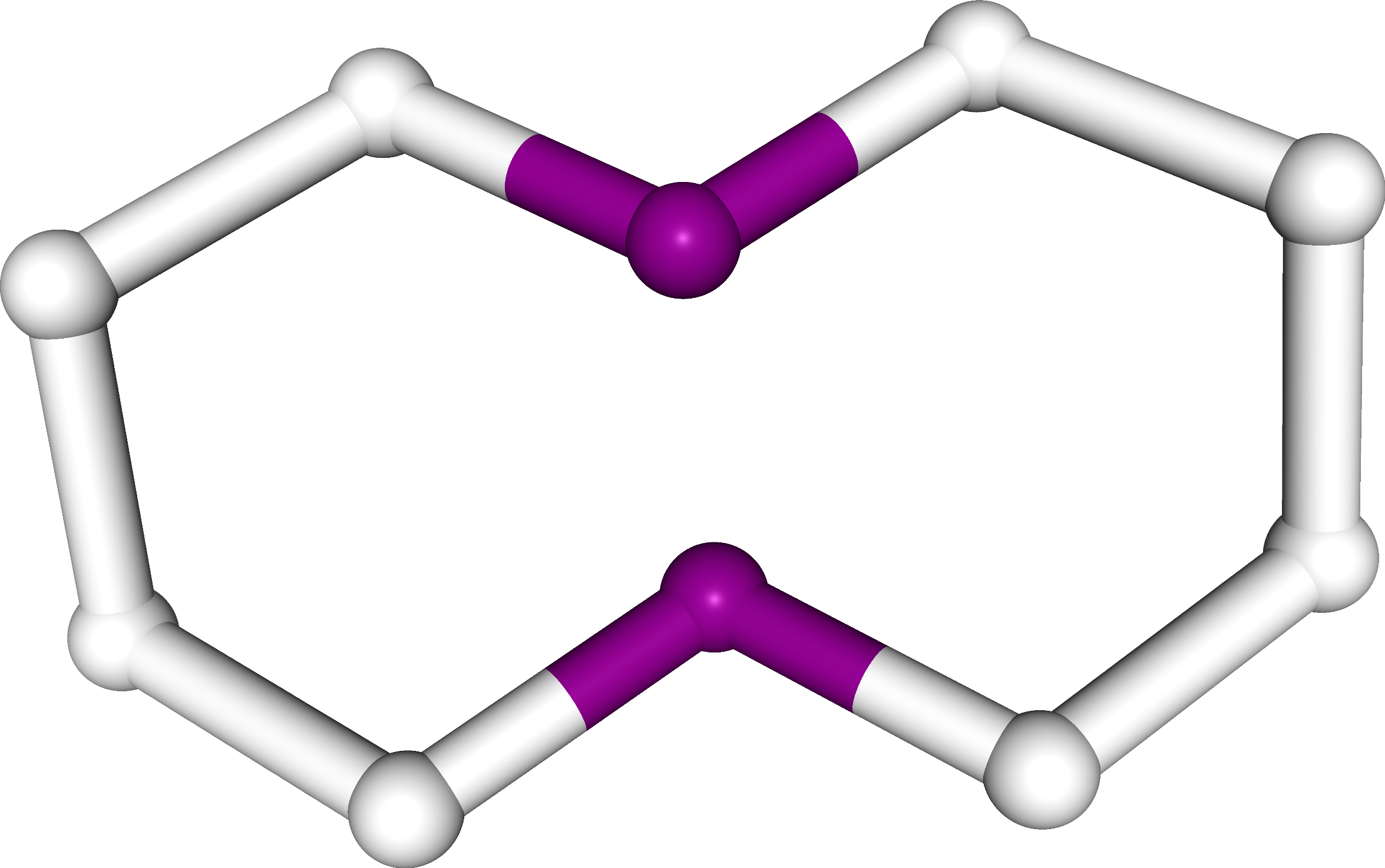}
    \caption{}
    \label{TB_tet}
  \end{subfigure}\hfill 
  \begin{subfigure}{.3\linewidth}
  \centering
    \includegraphics[width=\linewidth]{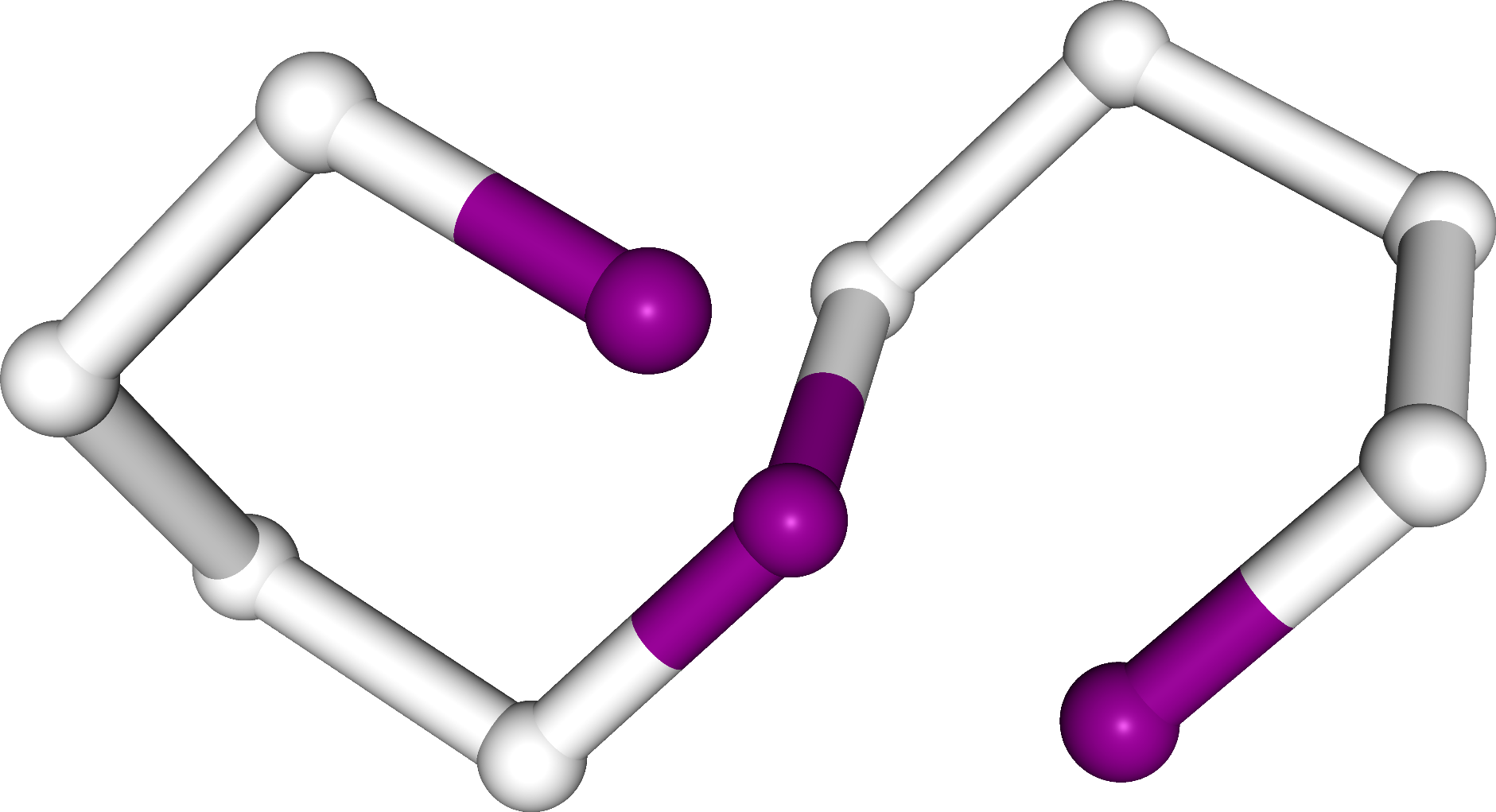}
    \caption{}
    \label{CB_tet}
  \end{subfigure}
  \hspace{1cm}
  \caption{Example of an unphysical ground state configuration obtained from the turn-based tetrahedral model next to the ideal physical configuration. The considered sequence is given by HPPPPHPPPPH in the HP-model. 
  (a) Unphysical lowest energy fold where beads 0 and 11 overlap. Since in the HP-model there is no interaction between H and P beads the chain can self-intersect without sacrificing energy. (b) Alternative ground state without overlap. Both folds have the same energy rendering them simultaneous ground states of the model.}
  \label{fig:UnphysicalConfig}
\end{figure}

\textcolor{\corrcolor}{Even though the model presents considerable improvements over the Cartesian grid,} we would like to highlight one issue: the model fails to adequately penalize overlaps, which can result in unphysical solutions within the feasible solution space. This can in some instances include the ground state leading to wrong folds. The root cause of this issue lies in the mathematical structure of the encoding.
Part of the model's better performance comes from its treatment of amino acid overlaps. It incorporates the overlap penalty into the interaction energy function, meaning that overlaps are only penalized near an interaction (see Appendix~\ref{Appendix:TBtet})
\begin{equation}
  H_\text{int} = q_{ij} \left(\epsilon_{ij} + \lambda_1 (D(i,j) - 1) + \sum_{r \in \mathcal{N}(j)} \lambda_2 (2 - D(i,r)) + \sum_{m \in \mathcal{N}(i)} \lambda_2 (2 - D(m,j)) \right),
\end{equation}
where $D(i,j)$ is the distance function between beads $i$ and $j$, the first Lagrangian multiplier $\lambda_1$ ensures that the interacting beads are nearest neighbors and the second multiplier $\lambda_2$ ensures that the neighboring beads are at distance $2$ on the grid.
In this formulation, the overlap is penalized only when two amino acids are close to a contact, and it is not penalized otherwise. 
While this approach scales much better than penalizing all possible overlaps, it has a significant drawback: the penalty is controlled by the interaction qubit \(q_{ij}\). The main issue with this form of penalization is that by turning off the interaction qubit (e.g. setting $q_{ij}=0$), the penalty can be completely avoided.
Hence, by ``sacrificing'' one interaction energy $\epsilon_{ij}$, the peptide chain can overlap. 
In most cases, this isn't an issue, as it's typically more energetically favorable to find a configuration where the interaction energy is utilized. 
However, in some configurations, it may be more advantageous for the chain to self-cross and establish a better interaction later in the sequence.
We demonstrate the consequence of this on a minimal artificial example in Fig.~\ref{fig:UnphysicalConfig}.

\paragraph{Coordinate-based Cartesian/tetrahedral}
When performing the scaling analysis of the different models, we found that the coordinate-based model performed better than the turn-based ones. 
\textcolor{\corrcolor}{The coordinate-based approach appears to be the most promising for quantum annealing. The native 2-local problem formulation enables an efficient representation on current-gen quantum annealers and does not require introducing additional qubits for locality reduction. Apart from requiring more qubits, the locality reduction also increases the strength of the penalty terms, which in turn demands higher coupler resolutions, something the coordinate-based approach avoids entirely.
Finally, at the current stage of hardware development, dense models are more costly to embed. Because the tetrahedral grid yields a sparser interaction matrix and allows smaller grids, the coordinate-based approach on the tetrahedral grid stands out as the most promising for current and near-term quantum annealers.}

\textcolor{\corrcolor}{Even though the coordinate-based approach appears to be the most promising}, we found that the proposed models are still too dense to be efficiently embedded onto the annealer topology for peptide sizes beyond a proof-of-principle calculation of $\approx$5-20 amino acids. 
Moreover, although the QUBO matrix becomes sparser as sequence length or lattice size increases, the number of required couplings per qubit rises, indicating that embeddings get more complex with longer chains. 
Since minor-embedding remains the principal computational bottleneck, these results show that future quantum annealers must offer a hardware graph with much higher connectivity, such that embedding the models is possible.\\

In summary, we find that currently none of the models appear suitable for large-scale implementation on quantum annealers, although the coordinate-based models being more promising, however.
Each proposed model is limited, either by having overly dense QUBO matrices or by scaling issues, such as the increasing qubit connectivity required with longer peptide chains or large required coupler resolutions. 
\section{Quantum annealing vs. simulated annealing}
\label{sec:comp}
We now turn our attention to a performance comparison for the four different models using simulated annealing and, due to limited access to the D-Wave hardware, compare the scaling of quantum annealing for the most promising model (coordinate-based on a tetrahedral grid) with simulated annealing.

\subsubsection*{Dataset}
To perform the benchmark, we generate $100$ random instances of proteins for sequence lengths of $10$ residues, uniformly sampled from the $20$ naturally occurring amino acids.
To compare the scaling we consider subsections of increasing length ranging from $N = 4$ up to $N = 10$.
In contrast to Ref.~\cite{outeiral2021investigating}, we generate the sequences randomly without post-selecting those with a unique energy minimum. We make this choice because we do not intend to capture the expected behavior of real proteins, instead, we merely wish to compare the performance of the different formulations.

The estimation of the time-to-solution requires the ground state energy of each protein. The energy was determined via our implementation of the parallel tempering algorithm. 
While parallel tempering itself is a heuristic algorithm, it is extremely unlikely that lower energy states exist due to its fast convergence for these small problem instances. 
All PT simulations were performed with 400 temperatures for an increasing number of sweeps ranging from $10^1$ to $10^6$, where for most instances no new best configurations were found after approximately $10^3$ sweeps.

\subsubsection*{Time-To-Solution metric}
With the dataset defined, we shift our focus to investigate the performance of the models using a set of selected solvers. 
The comparison of the models is possible if they use the same lattice structure. Although the models differ in formulation, they encode the same problem and thus share the same ground state energy. 
We benchmark the problems according to a well-known performance metric used to compare quantum annealing with other heuristic solvers, called the time-to-solution (TTS).
The TTS defines the expected time, which the algorithm requires to find the ground state within a selected probability, usually chosen to be $99\%$. 
The TTS is calculated by multiplying the average runtime \( \tau \) for a single iteration of the algorithm by the expected number of runs
\begin{equation}
  \text{TTS} = \tau \cdot \frac{\log(1-0.99)}{\log(1-p_\text{ground})}.
\end{equation}

As has been stated in different works~\cite{katzgraber2015seeking, ronnow2014defining}, the TTS suffers from one major drawback. Generally, there is a trade-off between increasing the probability of finding the ground state by extending the search time and increasing the total number of runs while utilizing shorter individual run times per search. This leads to the issue that an observed scaling advantage can be misleading if the success probability is too high for a given problem. To alleviate this issue the TTS needs to be optimized for each data point.

\subsubsection*{Simulated annealing}
As a baseline heuristic to compare with, we investigate the scaling of the models using our in-house GPU-accelerated simulated annealing implementation. 
To this end we compare the performance of the generated data set between the turn-based and coordinate-based models. 
Supplementary information regarding the optimized cooling rate can be found in Appendix~\ref{Appendix:SA}.

\begin{figure}[!htb]

  \includegraphics[width=\linewidth]{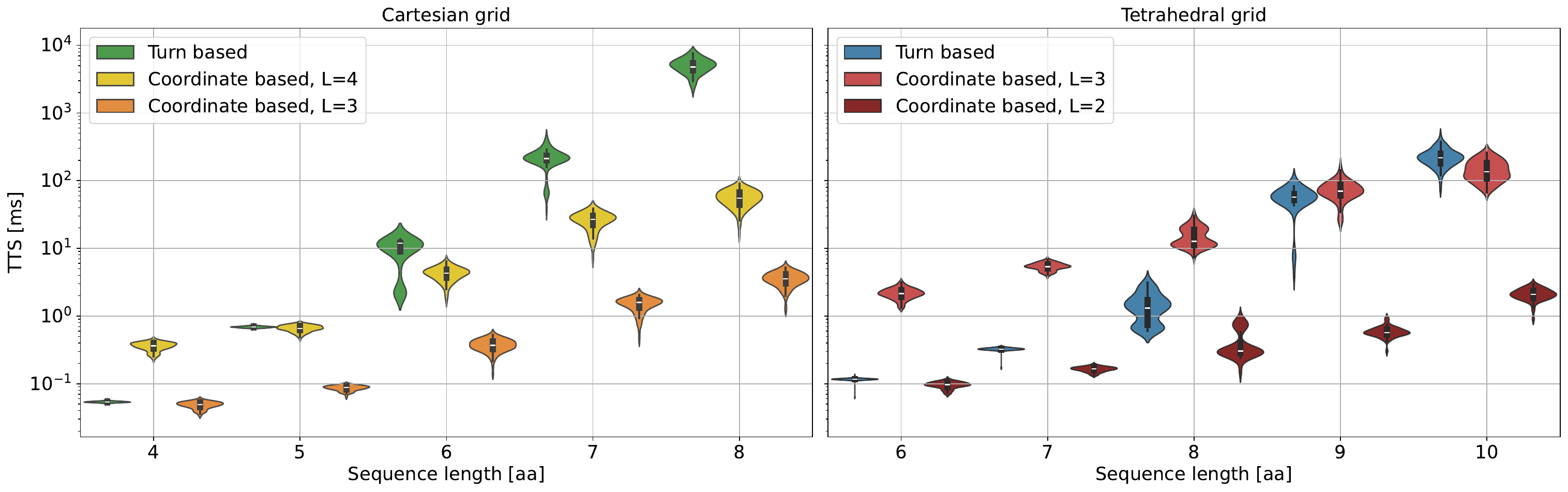}
  \caption{TTS scaling of the proposed models under simulated annealing. The data is taken over 100 randomly generated amino acid sequences. Results are shown for a Cartesian grid (left panel) and a tetrahedral grid (right panel). Due to free choice of lattice sizes the coordinate-based models have been evaluated on the minimal lattice size, such that the ground state still fits on the grid, as well as one size above this size.}
  \label{fig:TTS_SA}
\end{figure}

The results for the performance of the models for simulated annealing are presented in Fig.~\ref{fig:TTS_SA}. \textcolor{\corrcolor}{Note for the runtime $\tau$ of a simulated annealing run, we only account for the time required for sampling and do not consider other additional timings (e.g., for the graph coloring) as this is negligible compared to the runtime of the SA heuristic itself.} 
For visual clarity, we display the results for tetrahedral grids in the right panel and the results for Cartesian grids in the left panel.
\textcolor{\corrcolor}{As described in Sec.~\ref{sec:methods}, reducing the problem to 2-local interactions with the chosen penalty strength $\alpha$ has a substantial impact on algorithm performance. 
In contrast to the scaling analysis in Sec.~\ref{sec:scaling}, here we study the turn-based tetrahedral model with a heuristically fine-tuned penalty strength. 
Since we were not able to find the same fine tuning for the turn-based Cartesian model we relied on the methods of Ref.~\cite{babbush2014construction} for near-optimal penalty strengths. 
The specific parameter choices of $\alpha$ and penalty strengths are reported in more detail in Appendix~\ref{app:models} at the end of each section.}

To investigate the effect of the underlying lattice size of the coordinate-based models, we consider two different lattice sizes for both grids. 
Somewhat unsurprisingly, we find that the effect of a larger grid seems to result in a constant offset in the TTS making the problem more difficult to solve without changing the expected scaling.

The data indicates that the coordinate-based approach outperforms the turn-based approach for the TTS. Contrary to our expectation, this trend also holds for the turn-based model on the tetrahedral grid, even though it requires fewer qubits and has a less dense QUBO matrix. 
The most likely explanation for this effect is the significant disparity in the magnitudes of the QUBO matrix elements. 
At higher temperatures, the algorithm can easily traverse the energy barriers associated with the constraints. 
However, in this regime the temperature is too high for the interaction energies to play a crucial role in the folding process.
Our findings demonstrate that, in addition to resource requirements, the overall structure of the model exerts a significant influence on its performance.

\subsubsection*{Quantum annealing}
In the previous subsection, we analyzed the scaling of the proposed models for the classical simulated annealing algorithm. Here, we shift our focus to quantum annealing, specifically examining the scaling behavior of two generations of D-Wave quantum annealers: the \textit{Advantage 1} and the \textit{Advantage 2 prototype}.
As previously mentioned, the limited connectivity of quantum annealers requires embedding the problem onto the hardware graph. The two systems differ in their underlying connectivity, with the \textit{Advantage 1} using the \textit{Pegasus} and the \textit{Advantage 2 prototype} using the \textit{Zephyr} architecture. To account for these differences, 1000 separate embeddings were computed per sequence length for each architecture.
As discussed in Sec.~\ref{sec:emb}, embeddings can be reused. Therefore, for all peptides of a given sequence length $N$, the embedding with minimal number of qubits was selected.

\begin{figure}[!htb]
  \includegraphics[width=\linewidth]{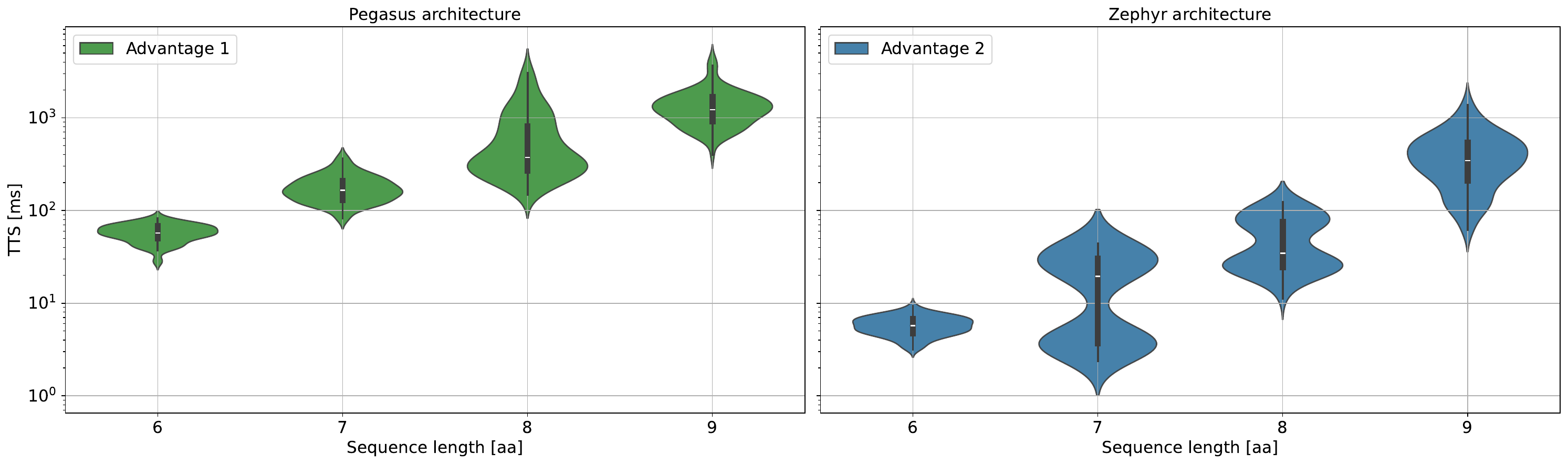}
  \caption{TTS scaling for the two tested quantum annealers: \textit{Advantage 1} (left panel) and \textit{Advantage 2 prototype} (right panel). The data shows the expected TTS for the coordinate-based model on the tetrahedral grid. Results indicate that the \textit{Advantage 2 prototype} achieves approximately an order of magnitude improvement over the \textit{Advantage 1}.}
  \label{fig:TTS_QA}
\end{figure}

To determine the optimal TTS for both systems, we performed an annealing time sweep ranging from \SI{1}{\micro \second} to \SI{1000}{\micro \second} for   \textit{Advantage 1} and from \SI{1}{\micro \second} to \SI{500}{\micro \second} for the \textit{Advantage 2 prototype} since we did not find substantial improvements beyond this range.
For both devices, the TTS decreases steeply up to approximately \SI{100}{\micro \second}, after which it plateaus. The optimal annealing times were found to be \SI{1000}{\micro \second} for the \textit{Advantage 1} and \SI{150}{\micro \second} for the \textit{Advantage 2 prototype}, explaining the order-of-magnitude advantage.
Additional details on the optimal annealing times are provided in Appendix~\ref{Appendix:QA}. \textcolor{\corrcolor}{Further details on the embeddings, including an analysis of the obtained chain length distribution and measured chain break frequencies are provided in Appendix~\ref{Appendix:Chainbreaks}. 
We want to highlight here that for the results in Fig.~\ref{fig:SA_comp_tet} we did not correct any chain breaks. 
In Appendix~\ref{Appendix:Chainbreaks} we quantify the effect of how much the TTS improves, if one uses majority voting to correct broken chains. we found that the effect is negligible, in particular for the scaling analysis.}

Figure~\ref{fig:TTS_QA} illustrates the TTS scaling for both devices, focusing on the most promising model identified, the coordinate-based model on the tetrahedral grid with sequence lengths ranging from $N = 6$ to $N = 9$. 
Both quantum annealers successfully solved all problem instances. Notably, the \textit{Advantage 2 prototype} outperformed the \textit{Advantage 1} by roughly an order of magnitude, underscoring the performance improvements between hardware generations.
However, it remains unclear whether this improvement is primarily due to the enhanced hardware connectivity, since the embeddings differ significantly in qubit requirements, or the reduction in error rates. Nevertheless, these results demonstrate the potential for further TTS reductions through future hardware advancements.
\subsubsection*{Comparison}
\begin{figure}
\centering
\includegraphics[width=\linewidth]{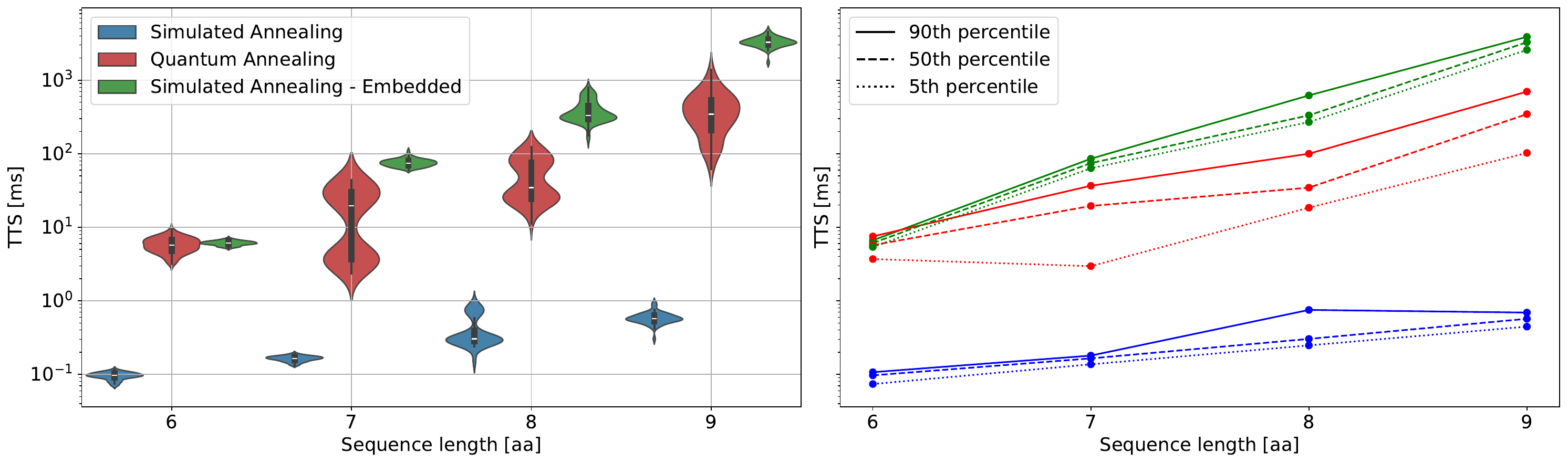}
\caption{Left panel: Scaling comparison of quantum annealing and simulated annealing. 
The blue curve shows the data obtained from simulated annealing on the problem before embedding it onto the annealer. 
The red curve shows the solution obtained from the Quantum annealer embedded on the Zephyr hardware graph. 
The green curve shows the results of simulated annealing on the embedded problem. 
Right panel: TTS scaling for the top 5\%, bottom 10\% and median percentiles. For the considered data, GPU-parallelized SA outperforms QA by several orders of magnitude. \textcolor{\corrcolor}{When considering the performance of the solution on the embedded problem, QA seems to outperform our in-house implementation of SA.}}
\label{fig:SA_comp_tet}
\end{figure}

Finally, we conclude with a direct performance comparison of QA and SA on the chosen model. 
The results for all considered sequences are presented in Fig.~\ref{fig:SA_comp_tet} as a violin plot with additional information regarding the 90th, 5th, and median percentiles.
We find that our GPU-parallelized implementation of SA significantly outperforms QA, with the performance offset being approximately proportional to the parallelization factor of $432$. 

\textcolor{\corrcolor}{While our analysis of the SOD in section~\ref{sec:spin_ovlp} showed that the models lie in a regime where quantum annealing due to tunneling could be advantageous, we did not observe a direct scaling advantage of quantum annealing over simulated annealing. However, it is important to note that the quantum annealer solves the problem after the embedding, which can be considerably harder to solve than the direct problem.}

To further assess the impact of the embedding, we evaluate SA performance also on the embedded problem. \textcolor{\corrcolor}{To this end we used the same QUBO matrix that was also solved by QA. 
The simulation results reveal that when the annealer solves the exact same problem as our SA implementation, QA outperforms our implementation of SA, leading to faster solutions even considering the parallelization speedup, as well as a possible scaling advantage for the sequence lengths considered in this work.}

Whether QA can achieve a speedup on the problem before the embedding remains to be seen in the future.
It is important to note that our results merely serve as an indication for the potential of quantum annealing and are by no means a rigorous scaling analysis. 
We finally address important caveats that need to be considered.

First, our results compare ``off-the-shelf'' versions of quantum as well as simulated annealing. This means that the tested algorithms do not utilize any prior knowledge of the problem, such as leveraging the one-hot-encoding structure for the placement of the amino acids, which could drastically speed up the computation time \cite{okada2019efficient}. 
Second, we did not consider any improvements to quantum annealing such as error correction schemes \cite{pudenz2014error} or the reverse annealing protocol \cite{chancellor2017modernizing}. 
Notably, Ref.~\cite{bauza2024scaling} was able to identify a scaling advantage for some optimization problems using error correction protocols. This highlights that reduced error rates can further improve solution quality as well as the scaling behavior.\\
\textcolor{\corrcolor}{Finally, our results are limited to very short peptide sequences. The considered sequence lengths in a range of $6-9$ amino acids are two short to draw reliable conclusions for the asymptotics. As sequence length grows beyond the current proof-of-concept, the exact scaling behavior remains uncertain. We expect the time-to-solution (TTS) to increase approximately exponentially with sequence length, although a super-exponential increase is not ruled out. While this growth is detrimental in principle, Ref.~\cite{outeiral2021investigating} notes that many clinically relevant proteins lie in the 300–1000 residue range, providing an empirical upper bound on the expected compute time. Even a modest quantum speedup could render such sequences accessible beyond what is achievable with classical computing.
}
\section{Conclusion and outlook}
\label{sec:conclusion}
We investigated and compared several of the proposed ab initio models to solve the coarse-grained protein folding problem on classical and quantum solvers. 
We evaluated these models in terms of their resource requirements, potential quantum advantage, and performance using simulated annealing and quantum annealing.
Our scaling investigation reveals that the coordinate-based approach seems more favorable for implementation on a quantum annealer, whereas the turn-based approach is limited by the locality reduction.

By performing the benchmark, we identified several issues, the most critical being the turn-based tetrahedral model from Robert \textit{et al.}~\cite{robert2021resource} producing unphysical configurations in the solution space. 
We further identified one more pressing bottleneck regarding all models: the number of qubit-qubit couplings required, which increases for all considered models with the sequence length. 
This number indicates how well a problem is suited for embedding onto an annealer's hardware graph, such as the \textit{Pegasus} or \textit{Zephyr} graphs. 
We found that for all considered models, this number increases, making it progressively more difficult to find embeddings as the number of amino acids in the protein increases. 
Another issue is the required coupler resolution of the turn-based models.
As the sequence length increases, the ratio between the largest and smallest coupling strength slowly increases. For larger sequences, this will necessitate an increasingly high coupler resolution, which is not supported by current-generation devices.

Additionally, we examined whether the proposed models are amenable to quantum speedup from tunneling by analyzing the spin overlap distribution, which serves as a proxy for the complexity of the free energy landscape. Our findings reveal that, to a large extent, the energy landscape is shaped by the problem encoding, particularly the constraints enforcing the qubits to represent a valid fold. 
While all models apart from the turn-based model on the Cartesian grid appear to operate in a regime where quantum speedup through the quantum tunneling effect is possible, we also observed that the embedding can significantly impact the spin overlap distribution.

Finally, we calculated the time-to-solution of simulated annealing for all models and compared with quantum annealing for the most promising one, the coordinate-based tetrahedral model.
In terms of scaling of SA, the coordinate-based model outperformed the turn-based models when expressed as QUBO problems. 
However, this advantage could shift in favor of turn-based models when formulated as HUBO problems. 
Our results show that simulated annealing and quantum annealing exhibit similar scaling behavior, but our GPU-parallelized implementation of simulated annealing outperforms quantum annealing by several orders of magnitude. 
Nevertheless, quantum annealing could, in principle, also be parallelized.
When comparing performance on the same problem, specifically the version embedded onto the quantum annealer, quantum annealing appears to scale better than our implementation of simulated annealing.

These findings indicate that, although there is currently no clear quantum advantage, quantum annealing could, in principle, achieve faster time-to-solutions than simulated annealing if the hardware can be improved, offering lower error rates and higher qubit connectivity.

\section*{Author Contributions}
TS and MH designed the project. AG implemented the simulated annealing and parallel tempering code. TS implemented the protein folding model and performed the experiments and numerical simulations. TS and MH performed the analysis. All authors contributed to the writing of this manuscript.

\section*{Acknowledgments}
We are grateful to Johannes Mueller-Roemer and Paul Haubenwallner for helpful discussions and comments on the manuscript. Furthermore, we would like to thank Philipp Quoss for valuable assistance with the implementation, which contributed to the technical aspects of this work.

\section*{Funding}
This work was supported by the research project Zentrum für Angewandtes Quantencomputing (ZAQC), which is funded by the Hessian Ministry for Digital
Strategy and Innovation and the Hessian Ministry of Higher Education, Research and the Arts.

\section*{Data availability}
All data regarding this publication are available in the corresponding GitLab repository:\\
\url{https://gitlab.cc-asp.fraunhofer.de/scheiber1/exploringquantumannealing4cgproteinfolding}

The repository contains all plotable data, the raw measurement results from the D-Wave devices as well as all QUBO matrices used in this study.

\subsection*{Code availability}
The underlying code regarding our implementations of simulated annealing or parallel tempering is not publicly available for proprietary reasons.

\subsection*{Competing interests}
The authors declare no competing interests.

\clearpage
\bibliographystyle{ieeetr} 
\bibliography{main}

\clearpage
\appendix
\section{Models}
\label{app:models}
In this appendix, we give a brief review of the PSP models considered in this work. The review here is by no means meant to be exhaustive. Additional details on the models can be found in the respective publications, which we cite at the start of each section. Further, we adjust some of the models to make them either more comparable or to reduce the number of qubits required while mapping to a two-local problem. While we aim to solve each model on a quantum annealer, it is important to note that some of these models were developed to be tackled with a gate-based quantum computer and might thus not perform optimally on a QA device.

Throughout this appendix, we will introduce the models in Boolean space. Although the variables are generic Boolean variables, we will refer to them as qubits $q$.

\subsection{Turn-based models}
Turn-based models encode the configuration of a protein by using coordinates relative to the origin of a coordinate system. The positions of the beads follow from a set of turns the polypeptide chain has taken. To prohibit the formation of unphysical configurations such as the chain folding back on itself or beads occupying the same lattice position, additional penalty terms have to be added.
The main advantage of turn-based models compared to coordinate-based ones is that the configuration can be stored in a linear amount of qubits.
However, to model interactions and formulate the penalties, additional variables need to be introduced.

\subsubsection{Cartesian lattice}
\label{Appendix:TBCart}
We start this section by presenting one of the first turn-based models, which was introduced in Ref.~\cite{perdomo2012finding} and refined in Refs.~\cite{babbush2014construction, babej2018coarse}.
Due to the bound locality of the QUBO matrix, a more efficient embedding on current QA devices can be performed.
The derivation in this appendix closely follows Ref.~\cite{babej2018coarse}, where the model has been considered on a three-dimensional Cartesian grid.

Turn-based models encode the folding of a protein as a self-avoiding walk by encoding the direction of a turn the amino acid chain takes. 
For example, a peptide chain on a three-dimensional Cartesian grid can grow in six possible directions, which must be encoded into qubits.

\begin{figure}[htp!]
    \centering
    \includegraphics[width=\linewidth]{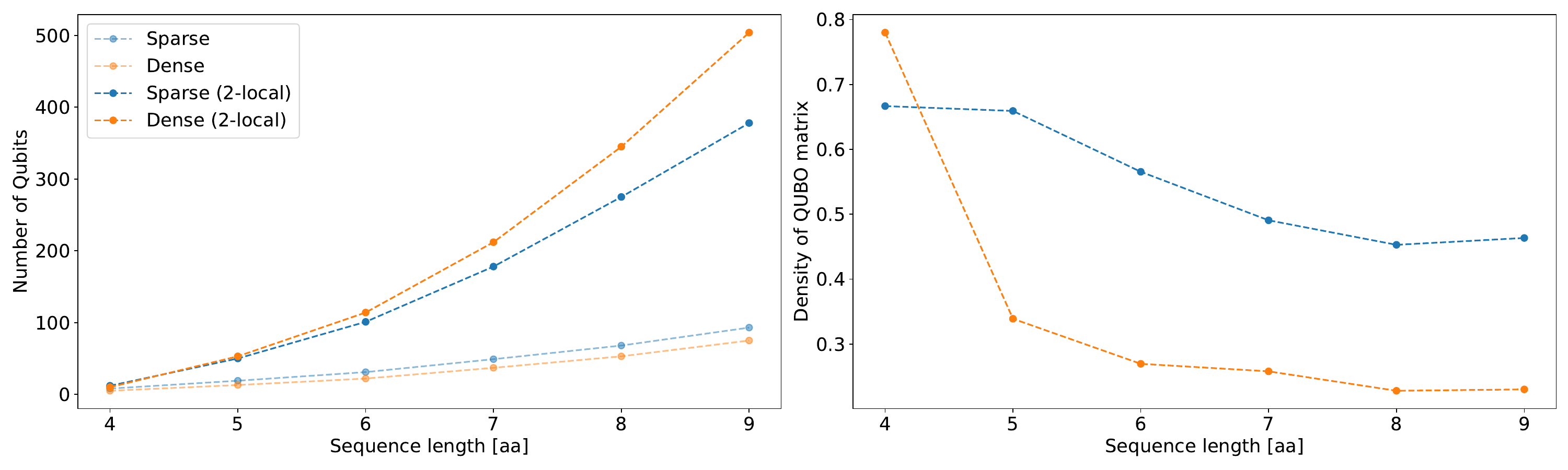}
    \caption{Scaling difference for the sparse and dense encoding. While the dense encoding leads to less qubits in the original problem it requires more qubits after the reduction to a 2-local model.}
    \label{fig:sparse_dense_cart}
\end{figure}

This encoding can be performed in two ways; either using a \textit{dense} or by using a \textit{sparse} encoding.
The dense encoding has the advantage of using fewer qubits: the possible directions a turn can take are directly encoded using binary variables.
Thus, encoding six possible spatial directions requires $\lceil \text{log}_2(6)\rceil = 3$ qubits. 

In this case the configuration of an amino acid is defined by the solution string
\begin{equation}
    \bold{q} = [101] [q_{4}01] \prod_{i=3}^{N-1} [q_{3i-2} q_{3i-1} q_{3i}].
\end{equation}

Due to symmetry reasons the first turn and most of the qubits from the second turn can be fixed. For mathematical simplicity we label the fixed qubits as if they were not restricted.

In the sparse encoding each direction is one-hot encoded and requires as many qubits as there are directions for each turn. In this case the configuration of an amino acid is defined by the solution string

\begin{equation}
    \bold{q} = [000001] [000q_{10}0q_{12}] \prod_{i=3}^{N-1} [q_{6i-5} q_{6i-4} q_{6i-3} q_{6i-2} q_{6i-1} q_{6i}].
\end{equation}

In this work we consider the dense encoding, since it leads to favorable performance. 
The reason for this can be seen in Fig.~\ref{fig:sparse_dense_cart}, which shows that while the sparse encoding requires a lower number of qubits, it leads to denser QUBO matrix and hence worse solver performance.

Since it is not possible to directly infer any information of the absolute position of the amino acids it is helpful to define turn-indicator functions. 
These boolean functions are used to evaluate in which direction a specific turn along the peptide chain has been taken and evaluate to \textit{True} if and only if the turn has been taken in the respective direction. 
In our case the indicators are given by
\begin{equation}
\begin{aligned}
    t^j_{+x} &= (1-q_{3j-2})q_{3j-1}q_{3j},  & t^j_{-x}  &= (1-q_{3j-1})q_{3j-2}q_{3j},  \\
    t^j_{+y} &= (1-q_{3j})(1-q_{3j-2})q_{3j-1},  & t^j_{-y} &= (1-q_{3j})(1-q_{3j-1})q_{3j-2},  \\
    t^j_{+z} &= q_{3j-2}q_{3j-1}q_{3j},  & t^j_{-z} &= (1-q_{3j-2})(1-q_{3j-1})q_{3j}, 
\end{aligned}
\end{equation}
where $t_{\pm x,y,z}^j$ evaluate to $0$ (\textit{False}) or $1$ (\textit{True}) and indicate if the turn $j$ has been taken in the positive or negative $x,y,z$-direction.
Furthermore, additional turn indicators can be defined for the two qubit configurations which do not encode a valid turn
\begin{equation}
\begin{aligned}
    t^j_{000}  &= (1-q_{3j-2})(1-q_{3j-1})(1-q_{3j}), \\
   t^j_{011}  &= (1-q_{3j-1})q_{3j-2}q_{3j}.
\end{aligned}
\end{equation}
To ensure that the configuration encodes a valid set of turns an energy penalty is introduced
\begin{equation}
    H_\text{turn} = \lambda_\text{turn} \sum_{i=1}^N (t_{000}^i + t_{011}^i)
\end{equation} 
which is only applied to ensure that qubits are not in one of the two states which do not encode a turn. 
To prohibit the peptide chain from folding back onto itself, an additional energy penalty is implemented utilizing the turn indicators as follows:
\begin{equation}
\begin{split}
    H_\text{back} =  \sum_{j=1}^{N}  &(t^j_{+x} \land t^{j+1}_{-x}) + (t^j_{-x} \land t^{j+1}_{+x})\\
+&(t^j_{+y} \land t^{j+1}_{-y}) + (t^j_{-y} \land t^{j+1}_{+y})\\
+&(t^j_{+z} \land t^{j+1}_{-z}) + (t^j_{-z} \land t^{j+1}_{+z}),
\end{split}
\end{equation}
where $\land$ denotes the logical AND, which is mapped to binary multiplication in the QUBO formulation.

Apart from penalizing back folding, the main reason to introduce turn indicators is to allow for the calculation of the absolute position of the amino acids, which is given by the sum of the number of positive and negative turns along an axis. 
For example the coordinates of the $m$-th amino acid is given by
\begin{equation}
    x_m = \sum_{j=1}^{m-1} (t^j_{+x} - t^j_{-x}),\quad
    y_m = \sum_{j=1}^{m-1} (t^j_{+y} - t^j_{-y}),\quad
    z_m = \sum_{j=1}^{m-1} (t^j_{+z} - t^j_{-z}),
\end{equation}
with the first amino acid occupying the origin $(0,0,0$).

From the positions we are able to calculate the distance between two amino acids, which we need to calculate the configuration energy. 
To avoid a square root in the calculation the squared distance
\begin{equation}
    D(j,k) = (x_k-x_j)^2 + (y_k-y_j)^2 + (z_k-z_j)^2
\end{equation}
is customary used.  

To penalize nonphysical overlaps of the protein, we have to ensure that $D(j,k) > 0$ for all possible pairs of beads $(j,k)$. 
To include this inequality in the optimization problem, it is transformed into an equality via the introduction of slack variables. 
First, it is important to notice that $0 < D(j,k) < (j-k)^2$, that is, the maximum distance between two beads is at most the square of all possible turns taken in one direction. 

To ensure that $D(j,k) > 0$, a slack variable $\alpha_{jk}$ is introduced for any pair ($i,j$), with
\begin{equation}
    0 \leq \alpha_{jk} \leq (j-k)^2 - 1.
\end{equation}
Using this definition of $\alpha_{jk}$ it follows that for all possible distances $D(j,k) > 0$ the equality 
\begin{equation}
\label{eq:tb}
    (j-k)^2 - D(j,k) - \alpha_{jk} = 0
\end{equation}
can be fulfilled for a specific integer value of $\alpha_{jk}$.

The realization of $\alpha_{jk}$ is made by introducing additional binary variables. 
The amount of additional variables needed can be calculated by considering the number of bits the binary representation of the maximum possible distance requires:
\begin{equation}
    \mu_{jk} = \lceil \text{log}_2((j-k)^2) \rceil \cdot((1+j-k) \text{ mod } 2).
\end{equation}
The second factor ensures that only additional variables are introduced if the beads are separated by an even number of turns, as beads separated by an odd number of turns cannot overlap by construction. 

From this definition, the slack variable $\alpha_{jk}$ can be defined as
\begin{equation}
    \alpha_{jk} = \sum_{l=0}^{\mu_{jk} - 1} q_l 2^{\mu_{jk} - 1 - l}.
\end{equation}
By constructing the slack variable $\alpha_{jk}$ is bounded by $0 \leq \alpha_{jk} \leq 2^{\mu_{jk}} - 1$, in contrast to the desired relation $0 \leq \alpha_{jk} \leq (j-k)^2 - 1$. 
Thus, to ensure that the equality can be satisfied, Eq.~\ref{eq:tb} needs to be restructured to
\begin{equation}
    2^{\mu_{jk}} - D(j,k) - \alpha_{jk} = 0.
\end{equation}
Finally, this allows us to formulate the overlap penalty term
\begin{equation}
    \gamma_{jk} = \lambda_\text{olap} \left(2^{\mu_{jk}} - D(j,k) - \alpha_{jk}\right)^2,
\end{equation}
where $\lambda_\text{olap}$ is a positive constant. Since the penalty needs to be applied to each pair of qubits that could possibly overlap, the full overlap Hamiltonian is given by:
\begin{equation}
    H_\text{olap} = \sum_{i=0}^{N-5} \sum_{j=i+4}^{N-1} \gamma_{ij} \cdot \left((1 + j - i) \mod 2\right),
\end{equation}
where the last factor ensures that an overlap penalty is applied only to beads that can possibly occupy the same lattice site.

Finally, the model needs to be able to assign correct interaction energies to adjacent amino acids. 
We consider only nearest-neighbor interactions, hence we wish to ensure that the interaction energy is only applied when beads are on adjacent lattice sites. 
To implement this, an additional interaction qubit $q_{jk}$ for each possible interaction is introduced. 
This qubit is in the state $\ket{1}$ if two amino acids interact and in the state $\ket{0}$ otherwise. 
The construction of the energy function is thus
\begin{equation}
    \theta_{jk} = q_{jk} \epsilon_{jk} (2 - D(j,k)),
\end{equation}
where $\epsilon_{jk}$ defines the interaction energy between the two amino acids. 
If the interaction energies $\epsilon_{jk}$ are chosen to be manifestly negative (as is the case for HP- and Miyazawa-Jernigan-type interactions), this formulation guarantees that for all distances $D(j,k) > 2$, the term becomes positive, so that by flipping the interaction qubit to the $\ket{0}$ state, no penalty is applied.
This interaction term is then applied to all interacting amino acids
\begin{equation}
    H_\text{int} = \sum_{j=0}^{N-4} \sum_{k=j+3} [(j-k) \mod 2] \, q_{jk} \epsilon_{jk} (2 - D(j,k)).
\end{equation}
It follows that the final Hamiltonian is given by
\begin{equation}
    H(\bold{q}) = \lambda_\text{back}\cdot H_\text{back} +\lambda_\text{turn} \cdot H_\text{turn}(\bold{q}) + \lambda_\text{olap}\cdot H_\text{olap}(\bold{q}) + H_\text{int}(\bold{q}).
\end{equation}
For all calculations considered in this work, we choose $\lambda_\text{back} = \lambda_\text{olap} = \lambda_\text{turn} = 20$, since we found that these penalties still led to correct results while being as small as possible.
Apart from the resource estimates, the reduction to 2-local was performed using the methods of Ref. \cite{babbush2013resource}.

\subsubsection{Tetrahedral lattice}
\label{Appendix:TBtet}
We now present the turn-based model on the tetrahedral grid following the derivation in Ref.~\cite{robert2021resource}, where it has been introduced for the first time. 
\textcolor{\corrcolor}{As stated in the main text, this model can produce unphysical states, which in some instances includes the ground state. We present one minimal example in Appendix~\ref{sec:unphysical_fold}.}
In its original formulation the considered model incorporates next$^n$-nearest neighbor interactions as well as a side-chain component.
To ensure comparability, we restrict this model to backbone folding and nearest neighbor interactions. 
In contrast to the Cartesian model, we chose the \textit{sparser} encoding introduced in Ref.~\cite{robert2021resource}. 

\begin{figure}[htp!]
    \centering
    \includegraphics[width=\linewidth]{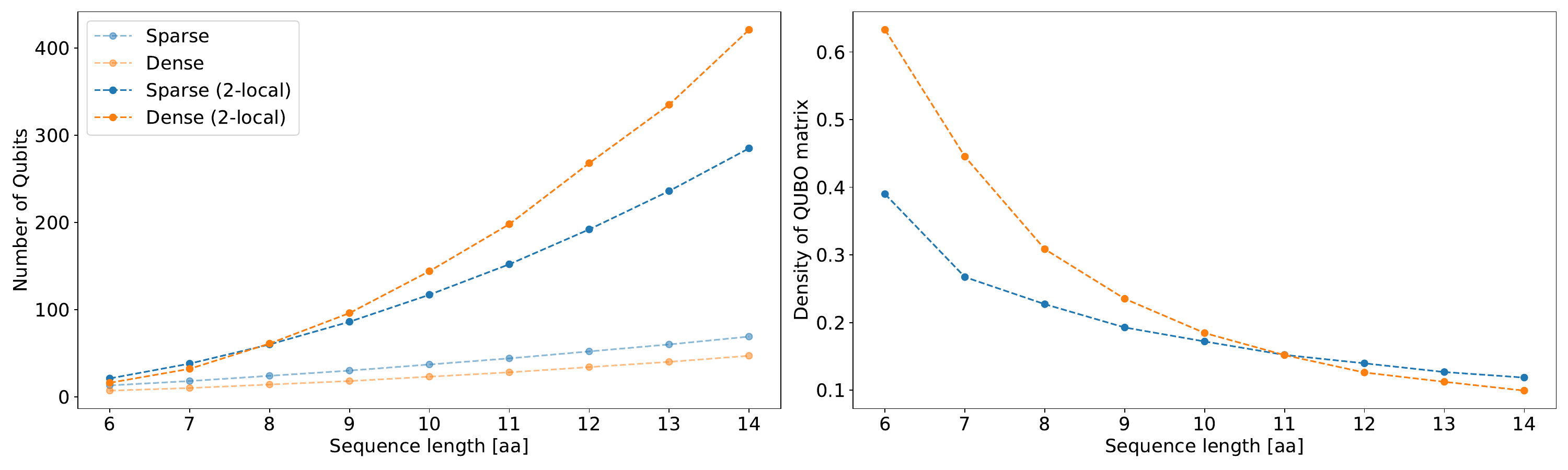}
    \caption{Scaling difference for the sparse and dense encoding. While the dense encoding leads to fewer qubits in the original problem it requires more qubits after the reduction to a 2-local model.}
    \label{fig:sparse_dense}
\end{figure}

We make this choice as for this model the sparser encoding leads to a lower number of qubits as well as sparser QUBO matrices as presented in Fig.~\ref{fig:sparse_dense}.
In the sparser, one-hot encoding, a turn is represented by one qubit for each possible turn the polypeptide chain takes. 
For the tetrahedral model, this corresponds to four possible directions:
\begin{equation}
   \bold{q} = [0001] [0010] \prod_{i=3}^{N-1} [q_{4i-3} q_{4i-2} q_{4i-1} q_{4i}].
\end{equation}
Due to symmetry reasons, the first two turns can be fixed, leading to some resource reduction. 
To infer the positions of the amino acids, it is again practical to introduce turn indicators.  
For the chosen encoding, these indicators are given by

\begin{equation}
t_0(i) = q_{4i-3}, \quad t_1(i) = q_{4i-2} \quad
t_2(i) = q_{4i-1},  \quad t_3(i) = q_{4i}. 
\end{equation}

With these turn indicators defined, it is possible to calculate the distance between any two beads $i$ and $j$ by counting the number of turns separating the beads along the chain
\begin{equation}
    \Delta n_a (i,j) = \sum_{k=i}^j (-1)^k t_a(k),
\end{equation}
where the factor of $-1$ keeps track of whether the turn has been made originating from an even or odd lattice site. 
Finally, the total distance between two beads can be calculated by taking the sum of the squared distances over the four axes:
\begin{equation}
    D(i,j) = \sum_a \Delta n_a(i,j)^2.
\end{equation}

With these definitions, the penalty functions of the model can be defined. 
Since we chose the sparser one-hot encoding, we need to ensure that the qubits will be in a state that encodes a turn. 
To achieve this, the first penalty term $H_\text{turn}$ is introduced:
\begin{equation}
    H_\text{turn} = \sum_{i=3}^{N-1} \lambda_\text{turn} \left(q_{4i-3} + q_{4i-2} + q_{4i-1} + q_{4i} - 1\right)^2,
\end{equation}
which ensures that only one of the qubits remains in the state $\ket{1}$ and penalizes all states where more than 1 qubit in a one-hot block are in the $\ket{1}$ state.

To prohibit configurations that are unphysical, i.e., two beads occupying the same lattice site, turns that lead to two beads overlapping need to be penalized. 
One possibility for two beads to overlap is back folding. 
To penalize two consecutive turns from growing in opposite directions, an additional growth constraint penalty
\begin{equation}
    H_\text{gc} = \sum_{i=3}^{N-1} \lambda_\text{gc} \sum_{a \in \{1,2,3,4\}} t_a(i) \land t_a(i+1)
\end{equation}
is added to restrict the growth of the chain to only include turns without back folding.

Finally, the interaction Hamiltonian, responsible for ranking the folds and prohibiting overlap not occurring from back folding, is introduced. 
In the original model this Hamiltonian is decomposed as
\begin{equation}
    H_\text{int} = H_\text{int}^{(1)} + H_\text{int}^{(2)} + H_\text{int}^{(3)} + \dots
\end{equation}
where the terms $H_\text{int}^{(n)}$ correspond to $n$-th nearest neighbor interactions. 
In this work we restrict the investigation to only nearest neighbor interactions, hence $H_\text{int} = H_\text{int}^{(1)}$.

The nearest neighbor interaction between two beads applies the interaction energy $\epsilon_{ij}$ if and only if two beads are nearest neighbors on the lattice. 

For each pair of beads $i$ and $j$ as in the turn-based model on the tetrahedral grid, there exists one interaction qubit $q_{ij}$, which is in the state $\ket{1}$ if and only if the two beads are in contact. 
To ensure that the energy is only applied if the beads are in contact, an additional penalty $q_{ij} \lambda_1 (D(i,j) - 1)$ is added. 
The penalty term $\lambda_1 > \epsilon_{ij}$ ensures that the interaction qubit is only in the state $q_{ij} = \ket{1}$ if the term in the parentheses vanishes. 
Note that $D(i,j)$ cannot be equal to 0 for two beads that are separated by an odd number of turns.

A further task of the interaction Hamiltonian is to penalize configurations where overlaps between two beads occur. 
As stated in Ref.~\cite{robert2021resource}, an overlap can only occur in the vicinity of a nearest neighbor interaction. 
The penalization of overlaps is then applied in a form such that, if a contact between two beads is established, another penalty term is added to ensure that the beads before and after bead $i$ do not overlap with bead $j$, and the neighboring beads to bead $j$ do not overlap with bead $i$
\begin{equation}
    H_\text{int}^{(1)} = \sum_{i=1}^{N-4} \sum_{\substack{j \geq i+5, \\ j-i=1 \mod 2}}^N h_{ij}^{(1)}
\end{equation}
with
\begin{equation}
\label{eq:En_Olap}
  h_{ij}^{(1)} = q_{ij} \left(\epsilon_{ij} + \lambda_1 (D(i,j) - 1) + \sum_{r \in \mathcal{N}(j)} \lambda_2 (2 - D(i,r)) + \sum_{m \in \mathcal{N}(i)} \lambda_2 (2 - D(m,j)) \right).
\end{equation}
To ensure that the term in parentheses in Eq.~\ref{eq:En_Olap} remains positive when the two beads, $i$ and $j$, are not in contact, the penalty terms must be chosen appropriately. Since we do not consider side-chains in our implementation, so it suffices to choose $\lambda_1 >  4 (j - i -1) \lambda_2 + \epsilon_{ij}$. 
The full Hamiltonian of the system is then described by
\begin{equation}
    H(\bold{q})  = H_\text{int}(\bold{q})  +  H_\text{gc}(\bold{q})  + H_\text{turn}(\bold{q}). 
\end{equation}

For the simulations considered, we choose $\lambda_2 = 10$  and scale the values of $\lambda_\text{turn} = \lambda_\text{gc}\equiv\lambda_\text{global}$ as a global penalty strength $\lambda_\text{global}$ to improve the performance of the solvers. 
That is, we scale the coefficient size with the chain length.
We found that scaling the coefficients in this way leads to better performance than using constant factors.
Namely for the calculation of the SODs we select a scaling of $\lambda_\text{global} = 21 \cdot N^3$, which leads to correct penalization of configurations that violate penalty terms.
For the TTS experiments we consider a scaling of $\lambda_\text{global} = 21 \cdot N^2$, which leads to smaller coefficients of the QUBO matrix but does not lead to a correct penalization of longer sequences.
This scaling has the benefit that it allows us to approximately scale the parameter of Rosenberg's polynomial with the same magnitude as the penalty terms $\alpha = 1.1 \cdot \lambda_\text{global}$. We would like to stress that this improvement originates from the reduction to 2-local and we do not expect there to be any benefit of scaling the coefficients when working with the model in HUBO-form.

We found that the same procedure cannot be applied to the turn-based model on the Cartesian grid, hence we restrict this method to this model only.

\subsection{Coordinate-based model}
\label{Appendix:CoordinateBased}
Coordinate-based models describe the fold of a protein by finding a mapping of the positions of the individual amino acids onto bits/qubits. 
Since in a direct encoding the amino acids can take any position, unphysical configurations need to be eliminated from the feasible solution space via penalty terms. 
In the following, we give a brief summary of the coordinate-based models considered in this work.

\subsubsection{Cartesian lattice}
We review the coordinate-based model presented in Ref.~\cite{irback2022folding} on a Cartesian (chessboard) lattice. 
In the model, an amino acid sequence $P = (a_1, a_2, \dots, a_N)$ is placed on a lattice $\mathcal{L}$ consisting of either $L^2$ sites in the 2-dimensional or $L^3$ sites in the 3-dimensional case.

To encode a configuration, one qubit for each amino acid is introduced at each lattice site. 
The qubit is in the $\ket{1}$-state if and only if the specific amino acid is placed at the specified lattice site. 
The total number of variables thus amounts to $N \cdot L^2$ for a 2-dimensional grid or $N \cdot L^3$ for a 3-dimensional grid. 

By choosing the further simplification that, on the Cartesian lattice, amino acids with an even index are positioned at even lattice sites (and vice versa for odd amino acids), the total number of variables can be reduced to $\approx N/2 \cdot L^2$ (or $N/2 \cdot L^3$).

The encoding of the state of a fold then takes the following form
\begin{equation}
    \bold{q} = \prod_{n=1}^{N_\text{even}} \left[q^n_{1}q^n_{2} \dots q^n_{L_\text{total}/2}\right] \prod_{n'=1}^{N_\text{odd}} \left[q^{n'}_{,1}q^{n'}_{2} \dots q^{n'}_{L_\text{total}/2}\right],
\end{equation}
where the first product considers the beads on even lattice sites and the second product considers the beads on the odd lattice sites \textcolor{\corrcolor}{and $L_{\rm total}$ refers to the total number of lattice sides}.

Since the formulation allows for unphysical configurations, i.e., multiple occurrences of the same amino acid or multiple amino acids on the same lattice site, three additional penalty terms are added to the energy function $H(\bold{q})$
\begin{equation}
    H(\bold{q}) = H_\text{int}(\bold{q}) + \sum_{i=1}^3 \lambda_i H_i(\bold{q}),
\end{equation}
where $H_\text{int}$ is the interaction energy of the amino acids, and the terms $H_i$ are the three positive penalty terms with the factor $\lambda_i$ denoting the relative strength of the penalty. 

Each of these three penalty terms ensures a different constraint that a physical configuration needs to fulfill.
The first term
\begin{equation}
    H_1 = \sum_{a \in P_\text{even}} \left(\sum_{s \in \mathcal{L}_\text{even}} q_s^a -1 \right)^2 + \{\text{same for odd parity}\}
\end{equation}
penalizes each configuration where a bead is located on more or less than one lattice site. 
Here, the first sum runs over all amino acids in the chain whereas the second sum runs over all lattice sites in the lattice $\mathcal{L}$.
The second term 
\begin{equation}
    H_2 = \frac{1}{2} \sum_{a_i \neq a_j} \sum_{s\in \mathcal{L}_\text{even}} q_{s}^{a_i} q_s^{a_j} + \{\text{same for odd parity}\}
\end{equation}
is used to prohibit two different amino acids $a_i$ and $a_j$ from being placed on the same position.
Finally, the third term
\begin{equation}
    H_3 = \sum_{1 < i < N} \sum_{s \in \mathcal{L}_\text{even}} q_s^{a_i} \sum_{\substack{s' \in \mathcal{L}_\text{odd},\\ ||s-s'|| > 1}} q_{s'}^{a_{i+1}} + \{\text{same with odd/even parity interchanged}\},
\end{equation}
is introduced to ensure that all amino acids lie on a chain. 
The last sum runs over all lattice sites $s$ and $s'$ which are not nearest neighbors on the lattice $||s-s'|| > 1$.

Apart from the penalty terms, whenever two beads are nearest neighbors an interaction energy is applied
\begin{equation}
    H_\text{int} = \sum_{|i - j| > 1} \epsilon_{ij} \sum_{\langle s, s'\rangle} q_s^{a_i} q_{s'}^{a_j}.
\end{equation}

One direct positive aspect of the model is that the locality of the interactions is bounded by 2. 
Further, the penalty strengths $\lambda_i$ do not scale with $N$ allowing for a direct implementation on a quantum annealer without the need of consideration for properties such as coupler resolution.

For all simulations considered we choose $\lambda_\text{1} = 18.6$, $\lambda_\text{2} = 14.4$, $\lambda_\text{3} = 18.6$, which is a heuristic choice inspired from the parameters chosen in Ref.~\cite{irback2022folding} adapted to the 3-dimensional grid and the Miyazawa-Jernigan interaction matrix. 
We would like to note that the results could likely be further improved by fine tuning the parameters.

\subsubsection{Tetrahedral lattice}
\label{Appendix:TetCart}
\begin{figure*}
    \hspace{2em}
    \label{fig:lower_error}
    \begin{subfigure}[h]{0.4\linewidth}
    \includegraphics[width=\linewidth]{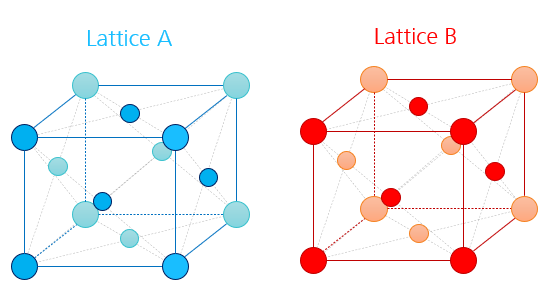}
    \caption{}
    \end{subfigure}
    \hfill
    \begin{subfigure}[h]{0.34\linewidth}
    \includegraphics[width=\linewidth]{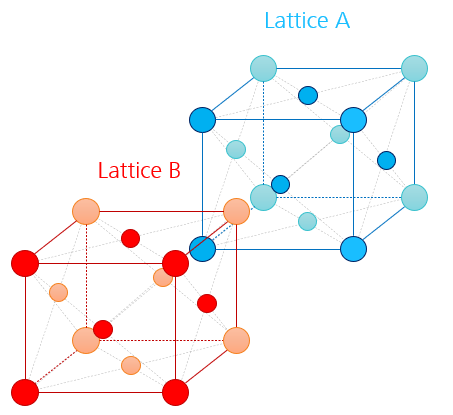}
    \caption{}
    \end{subfigure}%
    \hspace{2em}
    \caption{Example image of the two lattices forming the tetrahedral grid. Even beads are placed on lattice A while odd beads are placed on lattice B.}
\label{fig:tetrahedral_lattice}
\end{figure*}

To directly compare the model with the turn-based models presented earlier, we transition the coordinate-based model from a Cartesian lattice to a tetrahedral one.  
To this end, we propose a multi-grid implementation of the tetrahedral structure to adapt the coordinate-based model for this arrangement. 

Specifically, we define two interleaved face-centered cubic grids, $A$ and $B$, as illustrated in Fig.~\ref{fig:tetrahedral_lattice}. 
Each lattice is parametrized by three coordinates $(x, y, z)_{A,B}$ corresponding to the lattice vectors
\begin{equation}
    a_1 = (0, 1/2, 1/2), \quad a_2 = (1/2, 0, 1/2), \quad a_3 = (1/2, 1/2, 0)
\end{equation} 
for sub-lattice $A$ and the vectors
\begin{equation}
    b_1 = (1/4, 3/4, 3/4) \quad b_2 =  (3/4, 1/4, 3/4) \quad b_3 =  (3/4, 3/4, 1/4)
\end{equation}
for sub-lattice $B$.
Note that the lattices are related by a shift of a quarter diagonal.
From this, the coordinates of a bead can be derived as
 \begin{equation}
     (i,j,k)_A = i a_1 + j a_2 + k a_3
 \end{equation}
 for the first and 
 \begin{equation}
     (i,j,k)_B = i b_1 + j b_2 + k b_3
 \end{equation}
for the second sub-lattice.
Taken together, these sub-lattices form the tetrahedral structure, with vertices alternating between the two grids.

 With these definitions the PSP can be adapted from the coordinate-based model. 
 The amino acid sequence is again split into even and odd beads with the even beads living on sub-lattice $\mathcal{L}_A$, whereas the odd beads live on sub-lattice $\mathcal{L}_B$. 
 From this the penalty and interaction terms can be derived in a similar fashion as for the Cartesian grid:
 \begin{align}
     H_1 &= \sum_{a \in P_\text{even}} \left(\sum_{s \in \mathcal{L}_A} q_s^a -1 \right)^2 + \{\text{same for lattice $B$}\},\nonumber\\
    H_2 &= \frac{1}{2} \sum_{a_i \neq a_j} \sum_{s \in \mathcal{L}_A} q_s^{a_i} q_s^{a_j} + \{\text{same  for lattice $B$}\},\nonumber\\
    H_3 &= \sum_{1 < i < N} \sum_{s \in \mathcal{L}_A} q_s^a \sum_{\substack{s' \in \mathcal{L}_B,\\ ||s-s'|| > 1}} q_{s'}^{a_{i+1}} + \{\text{same with $A$/$B$ interchanged}\}.
\end{align}

Using this lattice split, it is important to define when two amino acids are adjacent to one another. 
On the tetrahedral grid, each lattice site generally has four nearest neighbor sites. 
For our model, we specify that for a grid position on the sub-lattice $A$ with coordinates $(i,j,k)_A$, the sites with coordinates $(i,j,k)_B$, $(i-1,j,k)_B$, $(i,j-1,k)_B$, and $(i,j,k-1)_B$ are nearest neighbor sites. 

With these definitions the interaction energy is given by
\begin{equation}
    H_\text{int} = \sum_{|i - j| > 1} \epsilon_{ij} \sum_{\langle s_a, s_b\rangle} q_{s_a}^{a_i} q_{s_b}^{a_j}.
\end{equation}

This approach represents a straightforward adaptation of the coordinate-based PSP onto the tetrahedral lattice, offering potential for broader applications. Our model employs a multi-grid implementation of the coordinate-based protein folding problem, allowing for generalization to more than two grids, which could further optimize resource utilization. Additionally, the multi-grid framework extends naturally to conjoint protein folding, where one protein is confined to one lattice and another to a separate lattice. This formulation enables efficient folding while inherently restricting folding domains and preventing overlap, making it applicable also to protein docking problems. 
The future potential of coordinate-based models remains open for exploration.

For all simulations considered we again choose $\lambda_\text{1} = 18.6$, $\lambda_\text{2} = 14.4$, $\lambda_\text{3} = 18.6$. It is likely that the performance can be further optimized if the parameters are fine-tuned.

\subsubsection{More efficient penalization}
The penalty term $H_3$ can be constructed in a more efficient form leading to a less dense QUBO. 
Hereby we do not penalize disconnected chain configurations but instead energetically favor configurations that are connected
\begin{equation}
    H_3 = (N-1) - \sum_{1 < i < N} \sum_{s \in \mathcal{L}_A} q_s^a \sum_{{\substack{s' \in \mathcal{L}_B,\\ ||s-s'|| = 1}} } q_{s'}^{a_{i+1}} + \{\text{same with $A$/$B$ interchanged}\}.
\end{equation}
Note that, to counteract the negative energy obtained by connecting two beads, we add a constant energy shift $N-1$ to the energy function. 
A similar approach has also been realized in Ref.~\cite{babbush2014construction}.
\section{Example of protein with unphysical fold on tetrahedral lattice}
\label{sec:unphysical_fold}

In this appendix we show how the result of obtaining unphysical folds with the turn-based tetrahedral lattice can be reproduced.
For this purpose, we utilized open-source code available in the Qiskit Community repository \cite{cb_tet_protein_folding}.
As discussed in Sec.~\ref{sec:disc}, the smallest protein for which we observe a non-physical fold as the ground state of the model consists of 11 amino acids and is represented in the HP model by the sequence HPPPPHPPPPH.
Since \cite{cb_tet_protein_folding} only supports MJ-type interactions, we consider the sequence LKKKKLKKKKL. This sequence mimics the behavior of the HP model, with strong interaction between the amino acids Leucine (L) - Leucine (L) and weaker interactions between the amino acids Lysine (K) - Lysine (K) and Lysine (K) - Leucine (L).

We calculated the Hamiltonian for this protein using the code mentioned above.
We find in total 8 different states with the same ground-state energy of $-1.474$.
The corresponding solution vectors are given by:
\begin{itemize}
    \item \ket{100000000001000110000100101} - correct
    \item \ket{100000000001001001001000101} - correct
    \item \ket{100000000001110110110100101} - correct
    \item \ket{100000000001111001111000101} - correct
    \item \ket{100000000001001011001000101} - wrong
    \item \ket{100000000001011011011000101} - wrong
    \item \ket{100000000001100111100100101} - wrong
    \item \ket{100000000001000111000100101} - wrong
\end{itemize}
out of which 50\% describe a correct fold and 50\% describe an unphysical configuration.

\section{Supplementary data for the simulations}

\subsection{Parallel Tempering}
\label{Appendix:Supp_ParallelTemp}





The parameters for the parallel tempering simulation for each experiment are depicted in Tab.~\ref{tab:PT_params}. 
The parameters include the number of temperatures, the minimum and maximum temperature and the number of sweeps.
The number of temperatures is chosen to be $400$ for all considered experiments as this is the number of replicas that can be run on the GPU without additional overhead influencing the run time. 
We chose a geometric spacing of the temperatures where the $i$th temperature is given by
\begin{equation}
    T_i = T_\text{min} \cdot r^{i}, \quad \text{where} \quad r = \left( \frac{T_\text{max}}{T_\text{min}}\right)^\frac{1}{(N_\text{temps}-1)}.
\end{equation}
This spacing is customarily used to ensure a higher density of temperatures on the lower end of the range and lower density of temperatures on the higher end.
For the calculation of the spin overlaps in Sec.~\ref{sec:spin_ovlp}, all PT runs were performed with a total number of $6 \cdot 10^6$.
However, to ensure thermal equilibration, only the last $10^6$ samples were used to produce the plots. 

To produce reference solutions for the dataset considered in Sec.~\ref{sec:comp}, we only ran PT simulations on the coordinate-based models, as the turn-based models have the same ground-state energy by design, with sweeps ranging from $10^1$ to $10^6$, as mentioned earlier.

\begin{table}[htp!]
    \centering
    \begin{tabular}{@{}lcc@{}}
    \toprule
    \textbf{Model} & 
    \textbf{T\_min} & 
    \textbf{T\_max}  \\ 
    \midrule
     Coordinate-based Cartesian & $1$ & $10^4$ \\
     Coordinate-based tetrahedral & $1$ & $10^4$ \\
     Turn-based Cartesian & $1$ & $10^{8}$ \\
     Turn-based tetrahedral & $1$ & $10^{6}$ \\
    \bottomrule
    \end{tabular}
    \caption{Parameters for the parallel tempering runs.}
    \label{tab:PT_params}
\end{table}

\subsection{Simulated Annealing}
\label{Appendix:SA}
\begin{figure}[htp!]
  \begin{subfigure}{0.5\textwidth}
    \includegraphics[width=\linewidth]{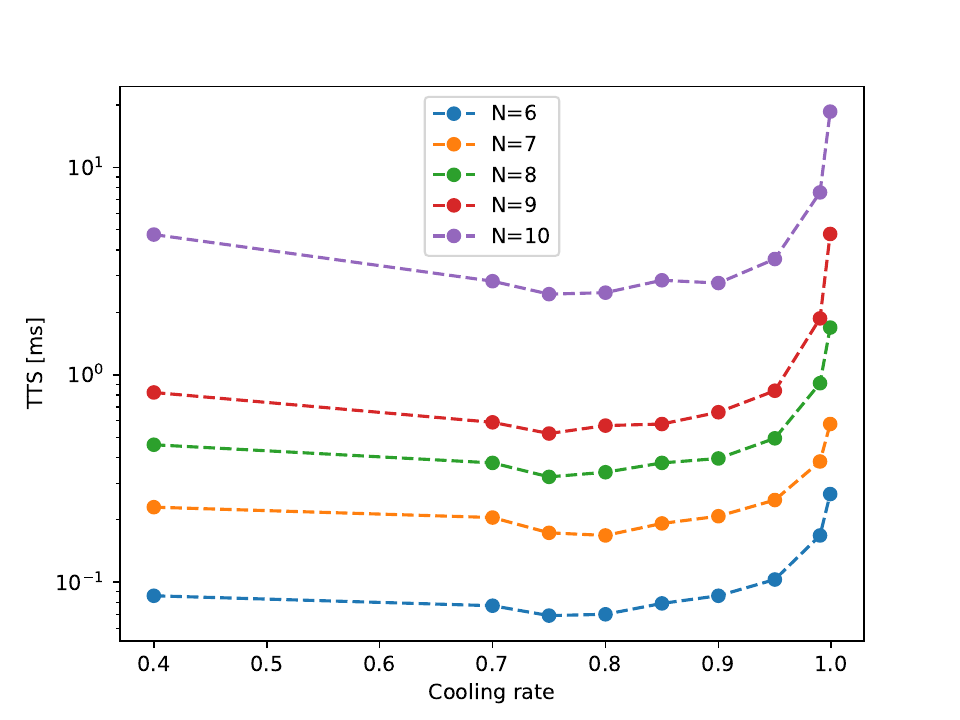}
    \caption{L = 2} \label{fig:1a}
  \end{subfigure}%
  \hspace*{\fill}   
  \begin{subfigure}{0.5\textwidth}
    \includegraphics[width=\linewidth]{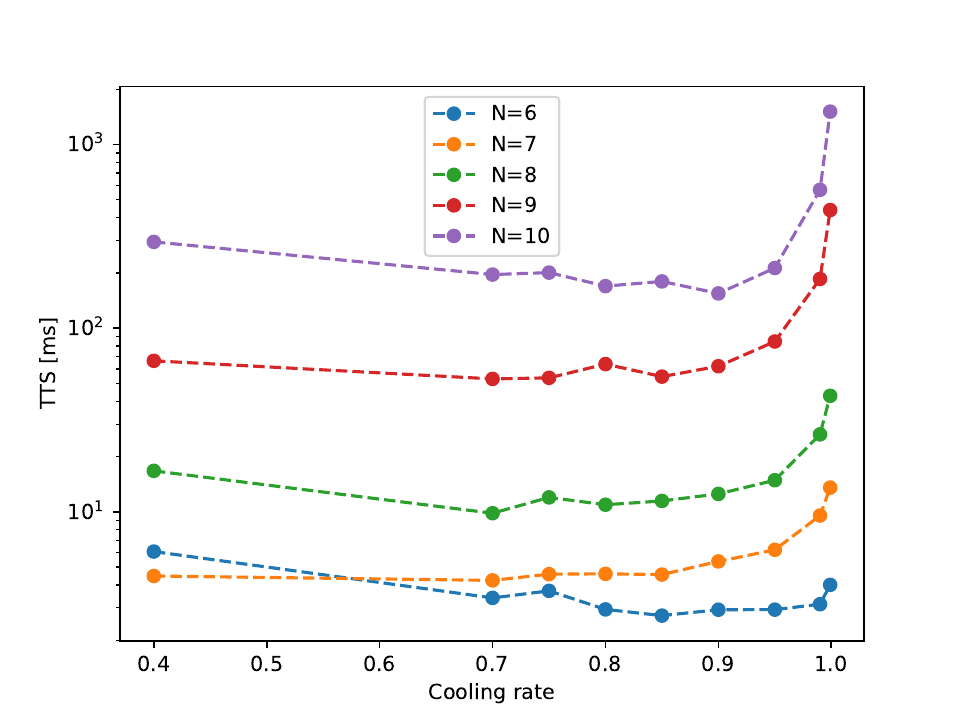}
    \caption{L = 3} \label{fig:1b}
  \end{subfigure}
\caption{Optimal cooling rate for the coordinate-based model on the tetrahedral grid for problem instances with varying amino acid sequence lengths $N$.} \label{fig:cooling_rate_tet}
\end{figure}

\begin{figure}[htp!]
  \begin{subfigure}{0.5\textwidth}
    \includegraphics[width=\linewidth]{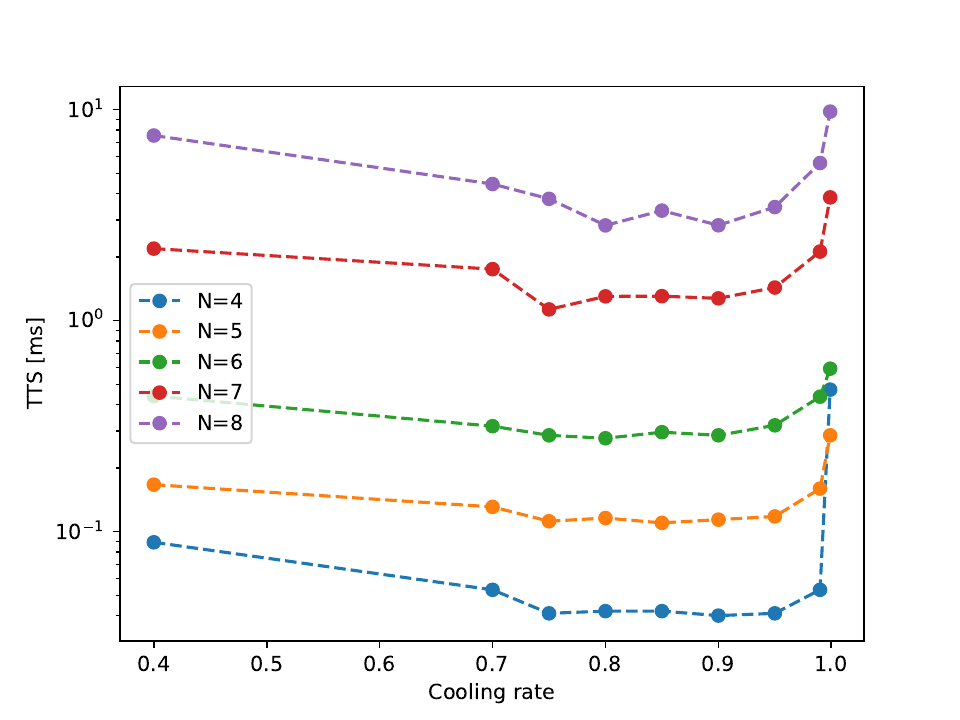}
    \caption{L = 3} \label{fig:1a}
  \end{subfigure}%
  \hspace*{\fill}   
  \begin{subfigure}{0.5\textwidth}
    \includegraphics[width=\linewidth]{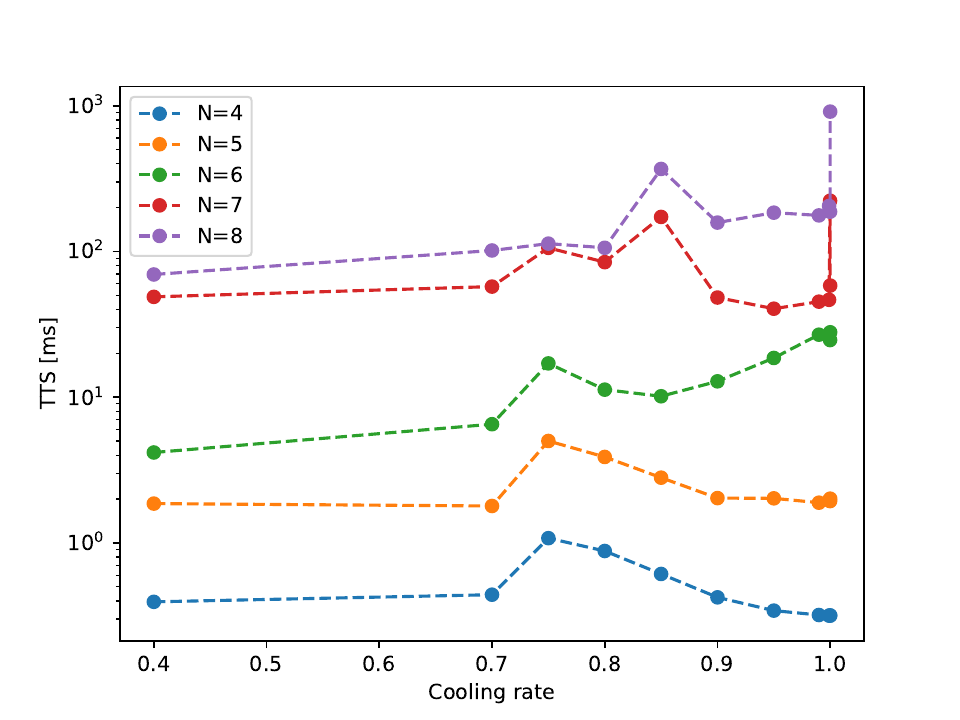}
    \caption{L = 4} \label{fig:1b}
  \end{subfigure}
\caption{Optimal cooling rate for the coordinate-based models on the Cartesian grid for problem instances with varying amino acid sequence lengths $N$.} 
\label{fig:cooling_rate_cart}
\end{figure}

\begin{figure}[htp!]
  \begin{subfigure}{0.5\textwidth}
    \includegraphics[width=\linewidth]{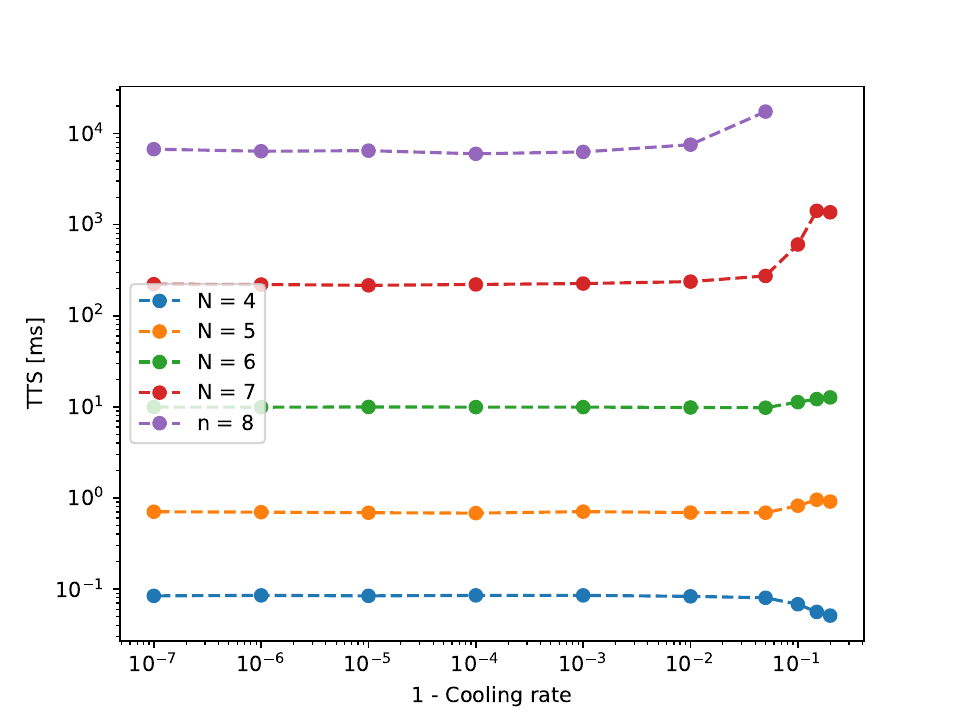}
    \caption{Cartesian} \label{fig:1a}
  \end{subfigure}%
  \hspace*{\fill}   
  \begin{subfigure}{0.5\textwidth}
    \includegraphics[width=\linewidth]{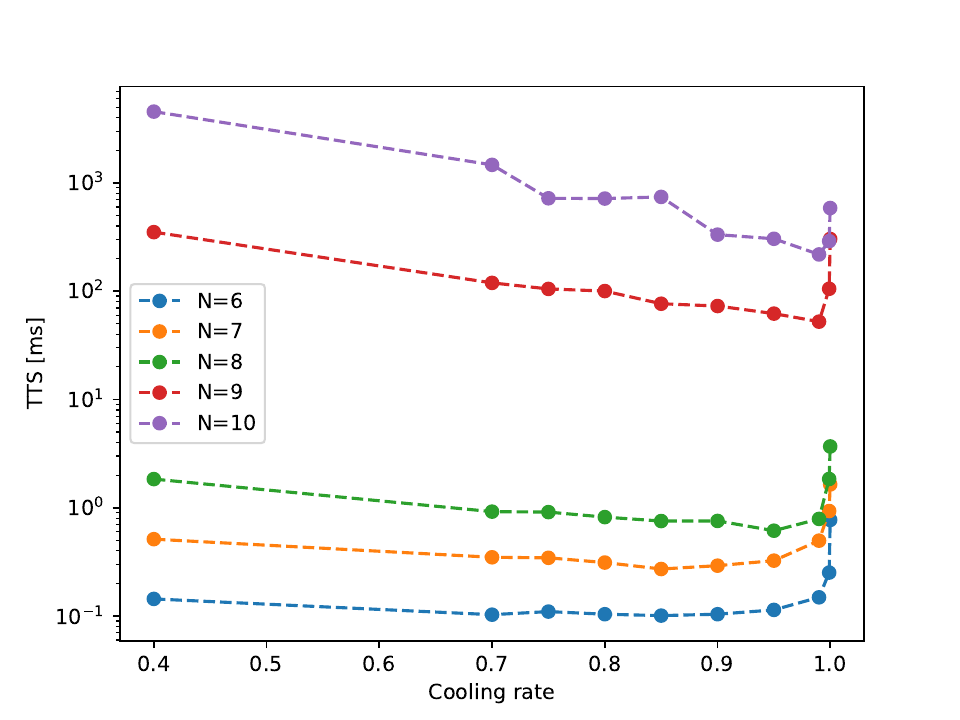}
    \caption{Tetrahedral} \label{fig:1b}
  \end{subfigure}
\caption{Optimal cooling rate for the turn-based models for problem instances with varying amino acid sequence lengths $N$. Since we had to choose cooling rates asymptotically close to 1, we plot 1 minus the cooling rate for the turn-based model on the Cartesian grid.}\label{fig:cooling_rate_tb}
\end{figure}

The only free parameter for the simulated annealing runs in our in-house implementation is the cooling rate $\zeta$, cf. Sec.~\ref{sec:SA}.
We optimized $\zeta$ for each problem instance by sweeping over different values.
The results are shown in Figs.~\ref{fig:cooling_rate_tet} (coordinate-based for the tetrahedral grid), \ref{fig:cooling_rate_cart} (coordinate-based for the Cartesian grid) and \ref{fig:cooling_rate_tb} (turn-based for tetrahedral and Cartesian grid).
For the numerical experiments in Sec.~\ref{sec:comp} we used the cooling rate leading to the minimal TTS for a given sequence length.

\newpage
\subsection{Quantum Annealing}
\label{Appendix:QA}
\begin{figure}[htp!]
    \centering
    \includegraphics[width=\linewidth]{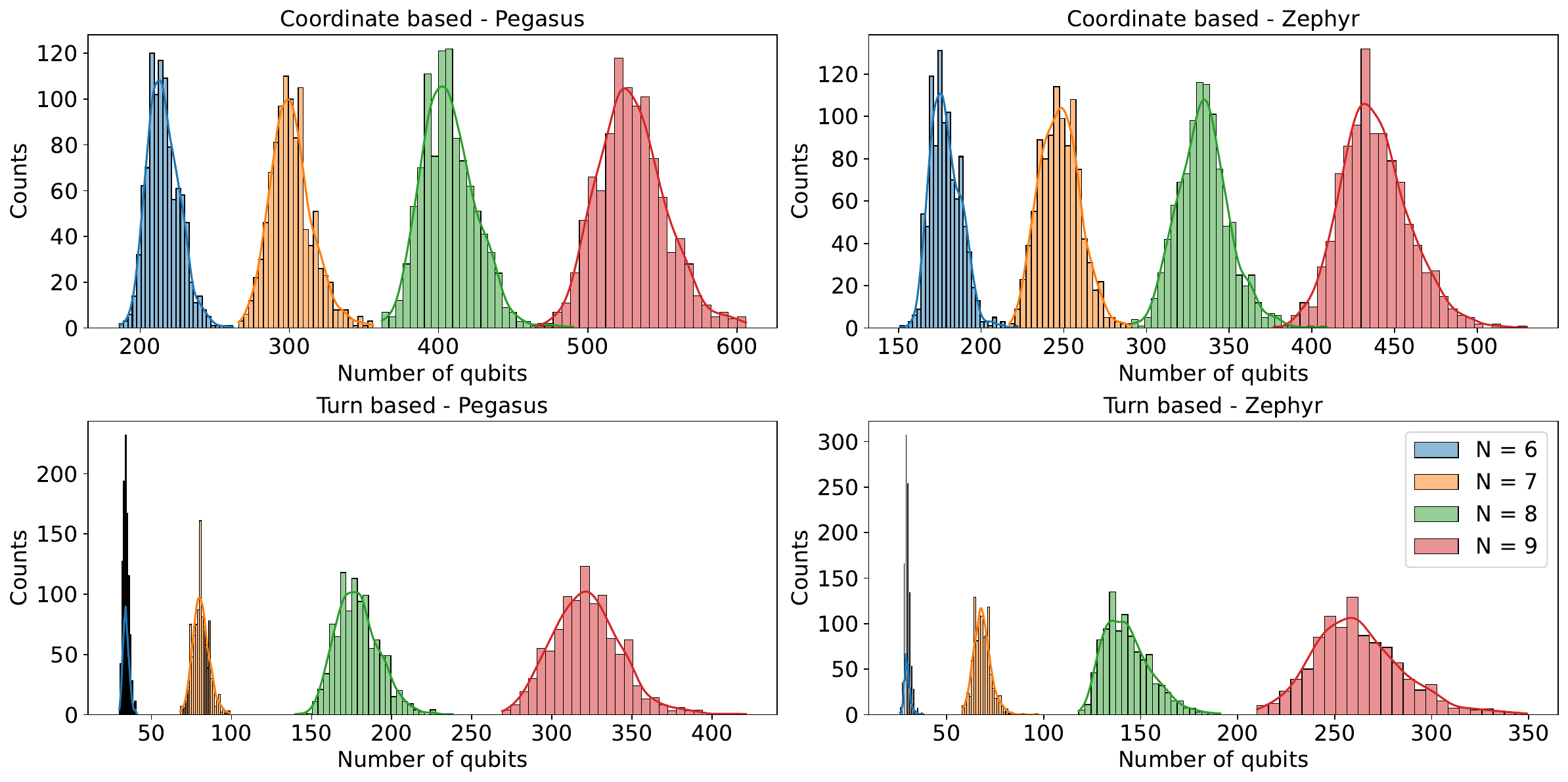}
    \caption{Embedding data for all models on the tetrahedral grid}
    \label{fig:embedding_tet_grid}
\end{figure}

\begin{figure}[htp!]
    \centering
    \includegraphics[width=\linewidth]{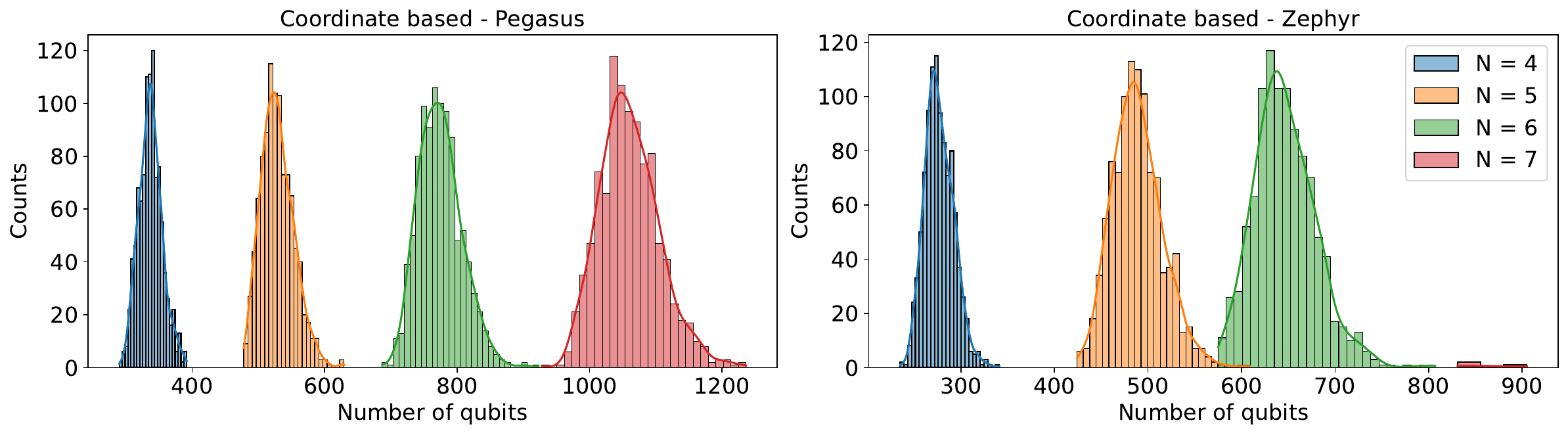}
    \caption{Embedding data for all models on the Cartesian grid. The last data point for $N = 7$ on the Zephyr graph shows only 3 data points. This is because only 3 out of 1000 conducted embedding processes returned a valid embedding on the prototype.}
    \label{fig:embedding_cart_grid}
\end{figure}

We first benchmark the embedding process for all considered models.
For each model we consider $1000$ embeddings and check the number of qubits needed on the \textit{Pegasus} or \textit{Zephyr} topology.
The results are shown in Fig.~\ref{fig:embedding_tet_grid} for the tetrahedral grid and Fig.~\ref{fig:embedding_cart_grid} for the Cartesian grid.

\begin{figure}[htp!]
  \begin{subfigure}{0.5\textwidth}
    \includegraphics[width=\linewidth]{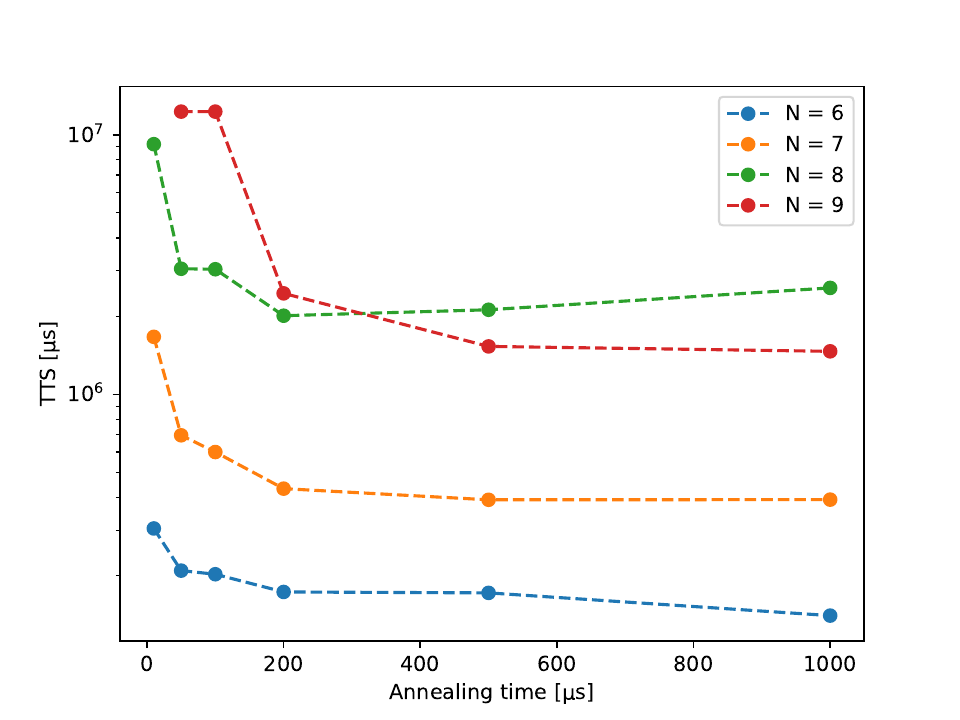}
    \caption{Advantage 1} \label{fig:1a}
  \end{subfigure}%
  \hspace*{\fill}   
  \begin{subfigure}{0.5\textwidth}
    \includegraphics[width=\linewidth]{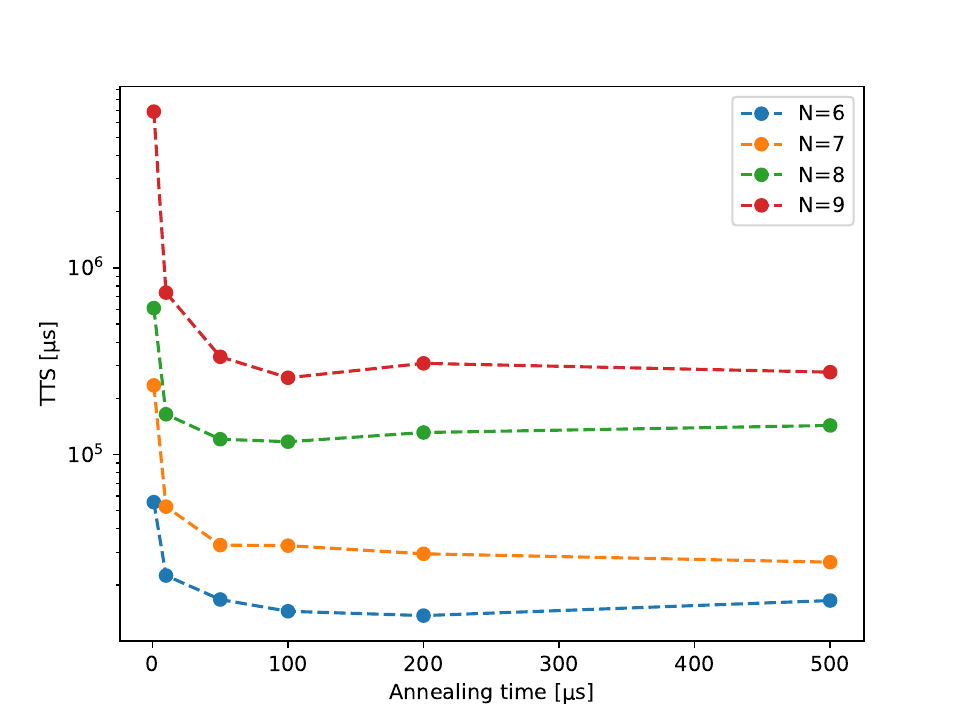}
    \caption{Advantage 2} \label{fig:1b}
  \end{subfigure}

\caption{\textcolor{\corrcolor}{Optimal anneal time to minimize the TTS for the D-Wave \textit{Advantage 1} and \textit{Advantage 2 prototype} systems. Since the anneal times show largely similar behavior, the optimal anneal times where chosen for all sequence lengths to be the same value of \SI{1000}{\micro\second} for \textit{Advantage 1} and \SI{150}{\micro\second} for the \textit{Advantage 2 prototype}.}} \label{fig:opt_anneal_time}
\end{figure}

Since the coordinate-based model on the tetrahedral grid turned out to be the most promising, we only run real hardware experiments for this model.
The relevant hyper-parameter to optimize the TTS for quantum annealing is the annealing time $t_a$ used for a given instance. 
We optimized $t_a$ for different instance sizes for the coordinate-based tetrahedral model.
The results for the \textit{Advantage 1} (based on the \textit{Pegasus} topology) as well as for the \textit{Advantage 2 prototype} (based on the \textit{Zephyr} topology) are shown in Fig.~\ref{fig:opt_anneal_time}.

\subsubsection{\textcolor{\corrcolor}{Chain break analysis}}
\label{Appendix:Chainbreaks}
\begin{figure}[htp!]
    \centering
    \includegraphics[width=\linewidth]{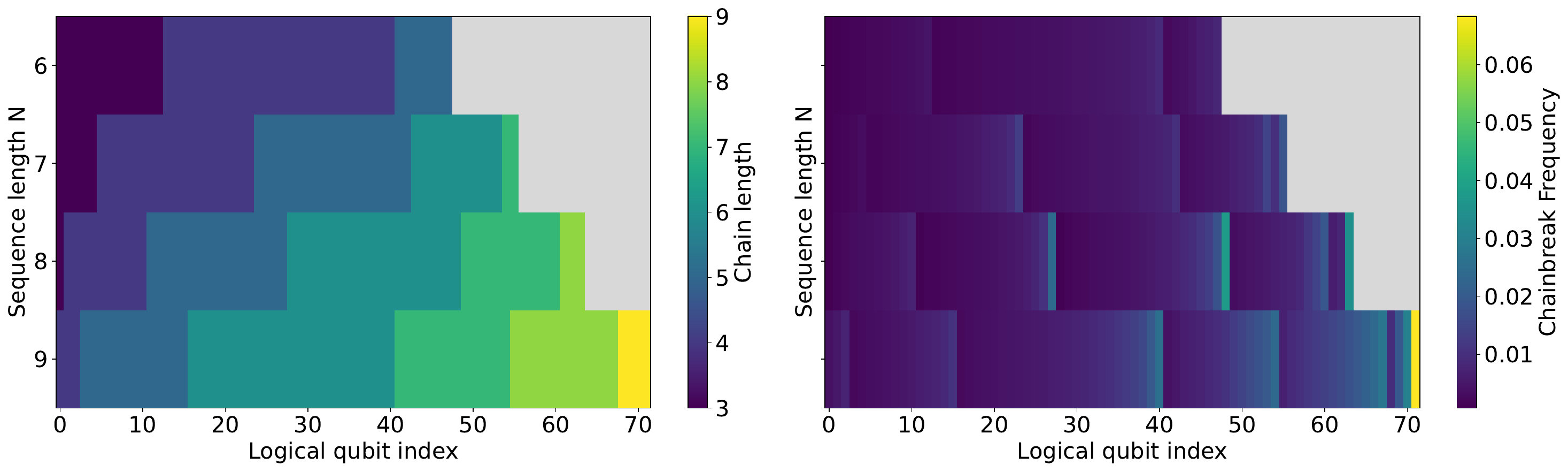}
    \caption{\textcolor{\corrcolor}{Distribution of the chain lengths of the embedded logical qubits (left)  as well as the observed chain break frequencies (right)  on the \textit{Advantage 1} annealer. The data is sorted by chain length as well as by chain break frequencies.}}
    \label{fig:ChainAdvantage1}
\end{figure}

\label{Appendix:Chainbreaks}
\begin{figure}[htp!]
    \centering
    \includegraphics[width=\linewidth]{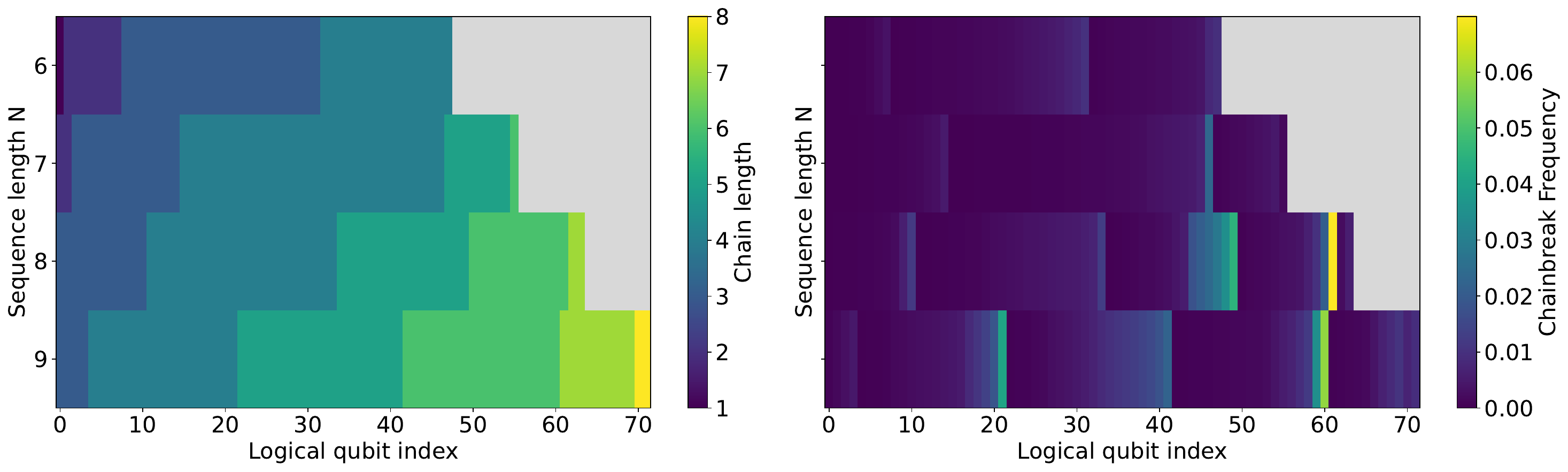}
    \caption{\textcolor{\corrcolor}{Distribution of the chain lengths of the embedded logical qubits (left) as well as the observed chain break frequencies (right) on the \textit{Advantage 2} annealer. The data is sorted by chain length as well as by chain break frequencies.}}
    \label{fig:ChainAdvantage2}
\end{figure}

\begin{figure}[htp!]
    \centering
    \includegraphics[width=\linewidth]{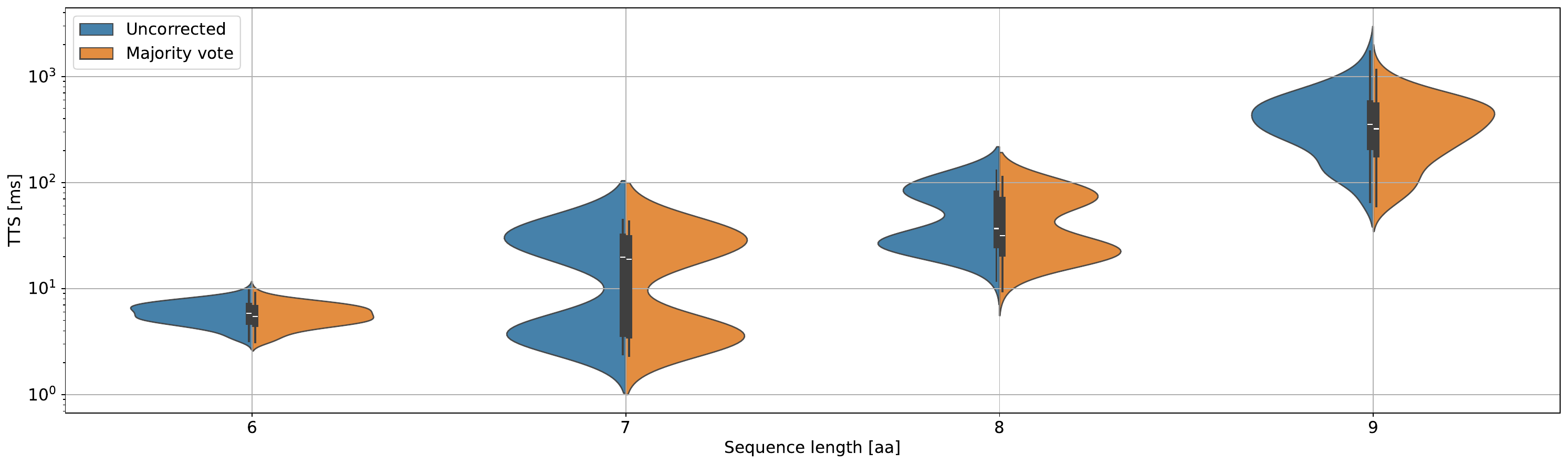}
    \caption{\textcolor{\corrcolor}{TTS improvements obtained by correction of chain breaks in the sample for the \textit{Advantage 2 prototype}. The left/blue part of the violin plot shows the obtained distribution before the majority vote, the right/orange part shows after the correction. While a slight improvement can be seen, the correction does not change the scaling in a meaningful way.}}
    \label{fig:ChainBreakCorrection}
\end{figure}

\textcolor{\corrcolor}{
Not all samples from the quantum annealing runs yield results corresponding to a correct embedding.
In this section, we provide a short analysis of chain breaks, including the distribution of chain lengths in the embedded problems as well as the observed chain break frequency. 
For each sequence length, the data is averaged over all sequences considered. 
The analysis is shown in Fig.~\ref{fig:ChainAdvantage1} for the \textit{Advantage 1} system and in Fig.~\ref{fig:ChainAdvantage2} for the \textit{Advantage 2} system. 
Our results show that some chains are substantially more prone to breaking than others, while the correlation with chain length is only mild. 
We attribute this to either degraded qubit performance at specific locations on the annealer or to more intricate dynamics of the specific qubits, which render these chains more likely to break.
Finally, in Fig. \ref{fig:ChainBreakCorrection} we show the influence of different chain break corrections on the TTS.
Specifically, we show that using a majority voting scheme, as it is provided in D-Wave Ocean SDK \cite{dwave_ocean}, leads to only an insignificant improvement in the TTS compared to the uncorrected results, where only results with unbroken chains are counted.
}

\subsubsection{Reliability of ground-state probability of the quantum annealer}
\begin{figure}
    \centering
    \includegraphics[width=\linewidth]{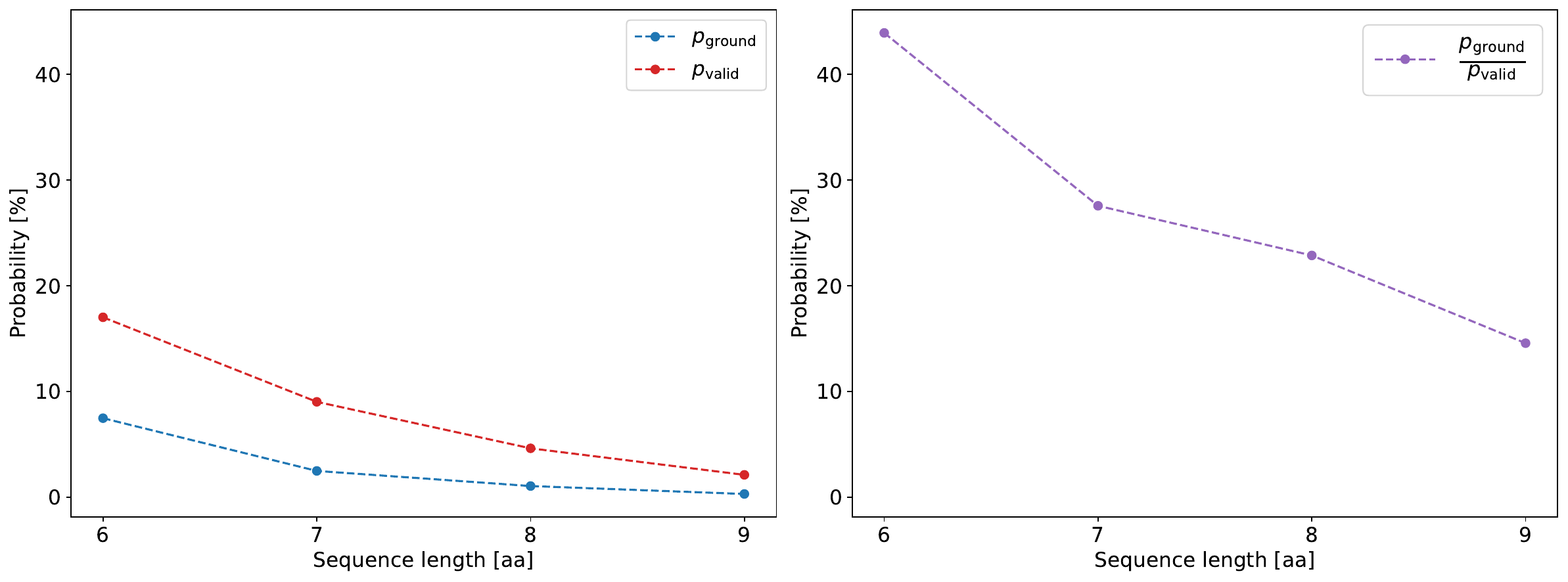}
    \caption{\textcolor{\corrcolor}{(Left panel) Relative probabilities of finding the ground state as well as finding a valid state for the \textit{Advantage 1} annealer. (Right panel) The number of ground states found per valid state. As expected this number decreases exponentially given the increase of the configurational space of the polypeptide.}}
    \label{fig:Reliability}
\end{figure}

\begin{figure}
    \centering
    \includegraphics[width=\linewidth]{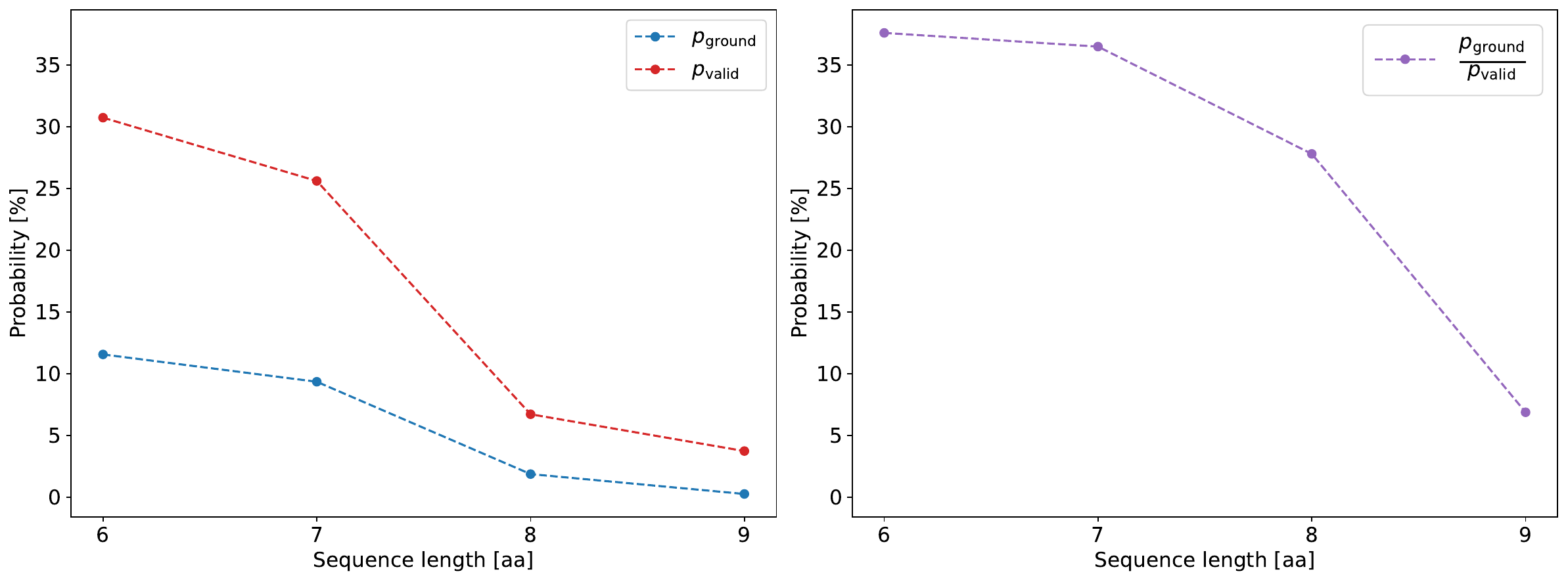}
    \caption{\textcolor{\corrcolor}{(Left panel) Relative probabilities of finding the ground state as well as finding a valid state for the \textit{Advantage 2 prototype}. (Right panel) The number of ground states found per valid state. As expected this number decreases exponentially given the increase of the configurational space of the polypeptide.}}
    \label{fig:Reliability_Pegasus}
\end{figure}

\textcolor{\corrcolor}{
To estimate how reliably the annealer is able to identify the correct ground state among the vast configurational space of the qubits of up to $2^{377}$, we estimated the probability of measured ground states as well as measured valid folds, that is solutions that do not violate the penalty terms. The results are shown in Fig.~\ref{fig:Reliability} and~\ref{fig:Reliability_Pegasus} and show that while initially the annealer can find the ground state reliably with a probability of over 10\%, for larger sequences this probability decreases steeply. This means that in order to estimate the results accurately either the ground state probability needs to be increased, by for example increasing the anneal time, or the number of shots needs to be scaled in order to resolve low ground state probabilities. 
Since for D-Wave quantum annealers the anneal time is upper-bounded by around \SI{2000}{\micro \second}, this means that for larger sequences the number of shots on the quantum annealer needs to be increased.
}

\end{document}